\RequirePackage{silence}
\documentclass[fleqn,usenatbib,useAMS]{mnras}

\usepackage{ae,aecompl}

\usepackage{graphicx}	
\usepackage{amsmath}	

\usepackage{amssymb}	
\usepackage{multicol}        
\usepackage{multirow}
\usepackage{bm}		
\usepackage{booktabs}
\usepackage{listings}
\usepackage{color}
\usepackage{threeparttable}
\usepackage{hyperref}
\usepackage{tikz}
\usetikzlibrary{calc}
\usetikzlibrary{decorations.markings}
\usetikzlibrary{shadings, calc, decorations.markings}
\usepackage{lineno}
\usepackage{ulem}

\defcitealias{Tremaine1984}{TW} 

\pubyear{2021} 
\begin{document} 


\title[Galaxy bar pattern speeds in $\Lambda$CDM and reality]{Fast galaxy bars continue to challenge standard cosmology}

\author[M. Roshan et al.]{\parbox[t]{\textwidth} {Mahmood Roshan$^{1,2}$\thanks{E-mail: \href{mailto:mroshan@um.ac.ir}{mroshan@um.ac.ir}}, Neda Ghafourian$^1$, Tahere Kashfi$^1$, Indranil Banik$^{3}$, Moritz Haslbauer$^{3, 4}$, Virginia Cuomo$^5$, Benoit Famaey$^6$ \& Pavel Kroupa$^{3, 7}$} \vspace{10pt} \\
$^{1}$Department of Physics, Faculty of Science, Ferdowsi University of Mashhad, P.O. Box 1436, Mashhad, Iran\\
$^{2}$School of Astronomy, Institute for Research in Fundamental Sciences (IPM), 19395-5531, Tehran, Iran\\
$^{3}$Helmholtz-Institut f\"ur Strahlen und Kernphysik (HISKP), University of Bonn, Nussallee 14$-$16, D-53115 Bonn, Germany\\
$^{4}$Max-Planck-Institut f\"ur Radioastronomie, Auf dem H\"ugel 69, D-53121 Bonn, Germany\\
$^{5}$Instituto de Astronom\'ia y Ciencias Planetarias, Universidad de Atacama, Copiap\'o, Chile\\
$^{6}$Universit\'e de Strasbourg, CNRS, Observatoire astronomique de Strasbourg, UMR 7550, F-67000 Strasbourg, France\\
$^{7}$Astronomical Institute, Faculty of Mathematics and Physics, Charles University, V Hole\v{s}ovi\v{c}k\'ach 2, CZ-180 00 Praha 8, Czech Republic}

\label{firstpage}
\pagerange{\pageref{firstpage}--\pageref{lastpage}}
\maketitle

\begin{abstract}

Many observed disc galaxies harbour a central bar. In the standard cosmological paradigm, galactic bars should be slowed down by dynamical friction from the dark matter halo. This friction depends on the galaxy's physical properties in a complex way, making it impossible to formulate analytically. Fortunately, cosmological hydrodynamical simulations provide an excellent statistical population of galaxies, letting us quantify how simulated galactic bars evolve within dark matter haloes. We measure bar strengths, lengths, and pattern speeds in barred galaxies in state-of-the-art cosmological hydrodynamical simulations of the IllustrisTNG and EAGLE projects, using techniques similar to those used observationally. We then compare our results with the largest available observational sample at redshift $z=0$. We show that the tension between these simulations and observations in the ratio of corotation radius to bar length is $12.62\sigma$ (TNG50), $13.56\sigma$ (TNG100), $2.94\sigma$ (EAGLE50), and $9.69\sigma$ (EAGLE100), revealing for the first time that the significant tension reported previously persists in the recently released TNG50. The lower statistical tension in EAGLE50 is actually caused by it only having 5 galaxies suitable for our analysis, but all four simulations give similar statistics for the bar pattern speed distribution. In addition, the fraction of disc galaxies with bars is similar between TNG50 and TNG100, though somewhat above EAGLE100. The simulated bar fraction and its trend with stellar mass both differ greatly from observations. These dramatic disagreements cast serious doubt on whether galaxies actually have massive cold dark matter haloes, with their associated dynamical friction acting on galactic bars.

\end{abstract}



\section{Introduction}
\label{Introduction}

The standard picture of galaxies residing within cold dark matter (CDM) haloes was historically motivated in part by the numerical result that self-gravitating Newtonian discs are very unstable to the formation of a very strong bar \citep{Miller_1968, Hockney_1969, Hohl_1971}. This can in principle be suppressed with a CDM halo as proposed by \citet{Ostriker_Peebles_1973}, but their equation 6 shows that they only considered a `rigid' halo which provides a fixed extra contribution to the potential, as would occur if e.g. the CDM particles in the halo did not move. If instead a live (responsive) halo is considered, the situation is more complex, as bars can be amplified \citep[e.g.][]{Sellwood_2016} or weakened \citep[e.g.][]{Bournaud_2005} through angular momentum exchange with the halo. In all cases, the dark halo rapidly slows down the bar pattern speed because it exerts efficient dynamical friction on the bar, as confirmed by both analytical investigations \citep{Tremaine1984} and $N$-body simulations \citep{Debattista2000}. The actual pattern speeds $\Omega_p$ of galactic bars can thus in principle provide an important test of the CDM picture.

Since galaxies come in a range of sizes, to know whether a bar is rotating fast or slow for its size, $\Omega_p$ is parametrized by the ratio of the corotation radius to the bar length. In other words, the important statistic is the dimensionless quantity
\begin{eqnarray}
    \mathcal{R} ~\equiv~ \frac{R_{\text{CR}}}{R_{\text{bar}}} \, ,
    \label{R_definition}
\end{eqnarray}
where $R_{\text{bar}}$ is the bar semi-major axis, and $R_{\text{CR}}$ is the bar corotation radius where the circular velocity $v_c$ plotted as a function of radius $R$ intersects a line with slope $\Omega_p$, i.e. these quantities satisfy the relation $\Omega_p R_{\text{CR}} \equiv v_c$.

If $1.0<\mathcal{R}<1.4$, then the bar is considered to be `fast' and extends almost until its corotation radius. But if $\mathcal{R}>1.4$, the bar is `slow' and corotation occurs far outside the bar radius. Ultrafast bars with $\mathcal{R}<1$ should not arise as such bars enter an unstable regime and dissolve \citep{Contopoulos1989}. Observations of bar pattern speeds in real galaxies firmly show that most bars are fast in the sense of having $\mathcal{R} \approx 1$ \citep{Corsini2011, Aguerri_2015, Cuomo2019b, Guo2019}.

In the CDM picture, however, dynamical friction tends to push $\mathcal{R}$ into the slow regime. The density profile of the dark halo plays a key role in the amount of friction \citep[e.g.][]{Read_2006}. Self-consistency with $\Lambda$CDM \citep{Efstathiou_1990, Ostriker_Steinhardt_1995} requires the CDM halo profiles and their velocity fields to be computed self-consistently in the theory by evolving small initial density perturbations into a population of galaxies 13.8 Gyr later. We therefore compare observed bars with those that form in the latest and highest-resolution $\Lambda$CDM galaxy formation simulations in a proper cosmological context. In particular, the EAGLE and IllustrisTNG simulations are used for this purpose. They rely on very different computational methods, which in principle makes the results robust. Our comparison with observations is done in terms of
\begin{enumerate}
	\item the fraction of galaxies that have bars, and
	\item the pattern speeds of the bars.
\end{enumerate}
A similar analysis for EAGLE previously revealed an $8\sigma$ tension with $\Lambda$CDM \citep{Roshan_2021}. The main advantage of the present work is a homogeneous analysis of several $\Lambda$CDM cosmological simulations using techniques similar to those applied to real galaxies, along with further improvements on the observational side \citep{Cuomo2020}.

The outline of this paper is as follows: In Section \ref{observed-OP}, we briefly review the current status of bar pattern speed observations, and describe the largest observational sample available to date. In Section \ref{cosim}, we introduce the cosmological simulations used in this paper, and discuss our selection rules to identify suitable barred galaxies. We also explain the techniques used to measure the bar strength and length, pattern speed, and corotation radius in the simulated galaxies. We present our results in Section \ref{res}. We then more carefully quantify the statistical distribution of $\mathcal{R}$ in simulations and in observations, thereby allowing us to find the level of tension between them (Section \ref{R_parameter}). Since the simulations overpredict $\mathcal{R}$, we consider a less accurate method to obtain the rotation curve that typically underestimates it, giving lower $\mathcal{R}$ and a more conservative estimate of the disagreement with observations (Section \ref{newCR}). We then discuss our results in Section \ref{Discussion}. Conclusions are drawn in Section \ref{Conclusions}.

\section{Bar pattern speed observations}
\label{observed-OP}

\citealt{Tremaine1984} (hereafter \citetalias{Tremaine1984}) developed a straightforward technique to recover the pattern speed $\Omega_p$ of a bar in a barred galaxy without involving complex dynamical modelling. With their approach, $\Omega_p$ is directly determined from observable quantities measured for a tracer population of stars or gas, which only has to satisfy the continuity equation. Further assumptions of the \citetalias{Tremaine1984} method are that the bar resides in a flat disc, and that its angular frequency is characterized by a well-defined $\Omega_p$. In its first application, the \citetalias{Tremaine1984} method was tested on a numerically simulated galaxy, resulting in a reliable measurement of $\Omega_p$ with 15\% accuracy.

Both photometric and spectroscopic data are needed to apply the \citetalias{Tremaine1984} method, which requires measurements of the surface density and line-of-sight (LOS) velocity distribution of the tracer along apertures located parallel to the major axis of the galactic disc. The first natural application of the \citetalias{Tremaine1984} method involved long-slit spectroscopy \citep[for a review, see][]{Corsini2011}. More than ten $\Omega_p$ estimates were obtained in mainly early-type galaxies, with typical uncertainties of around 30\%. In fact, the most suited tracer for the \citetalias{Tremaine1984} method is an old stellar population without contamination from gas.

The advent of integral field unit (IFU) spectroscopy allowed to reduce the uncertainties and enlarge the samples \citep[e.g.][]{Aguerri_2015, Guo2019}. IFU spectroscopy greatly reduces the main source of error in the \citetalias{Tremaine1984} method because it allows to define a posteriori the disc position angle (PA) along which to locate the apertures, to extract a sufficient number of apertures, and to correctly identify the centre of the galaxy and its systemic velocity. Thanks to these technological improvements, the method was applied to late-type galaxies as well, where dust and spurious elements (e.g. star-forming regions and/or other galaxy components like rings, warps, or strong spirals) may affect the surface mass density of the tracer. Nevertheless, the applicability of the \citetalias{Tremaine1984} method to spiral galaxies was tested by both theoretical \citep{gerssen2007, Zou2019} and observational studies \citep{Debattista_2004, Aguerri_2015, Cuomo2019b}. The assumption of a well-defined rigidly rotating pattern speed in a barred galaxy can be questioned in the presence of rings and/or spiral arms $-$ possible effects due to multiple or variable pattern speeds are unavoidable, but can be mitigated with a careful application of the \citetalias{Tremaine1984} method \citep{Debattista2002, Maciejewski2006, Meidt2008, Williams2021}.

Nowadays, the \citetalias{Tremaine1984} method can be applied using publicly available IFU surveys \citep{Cuomo2019b, Guo2019, Garma-Oehmichen2020}. Efforts have been made to apply the \citetalias{Tremaine1984} method using gaseous tracers as well \citep{Zimmer2004, Fathi2009}, but the reliability of the corresponding results is still under debate \citep{Williams2021}. Recently, the TW method can even be applied to stars in the Milky Way with a combination of Gaia and VVV proper motions \citep{Sanders2019}.

\citet{Cuomo2019} applied for the first time the \citetalias{Tremaine1984} method to data from the Multi-Unit Spectroscopic Explorer \citep[MUSE;][]{Bacon_2010}, reducing the uncertainties on $\Omega_p$ to $\la 10\%$. \citet{Cuomo2020} collected all previous results based on comparable applications of the \citetalias{Tremaine1984} method to investigate the relations between the properties of bars and their host galaxies. The authors built the largest sample of barred galaxies available to date with an available direct measurement of $\Omega_p$. The sample spans a wide range in morphological types (from SB0 to SBd), redshifts ($z < 0.08$), and Sloan Digital Sky Survey \citep[SDSS;][]{SDSS} $r$-band absolute magnitudes ($-23 < M_r < -18$). Most of the galaxies in the sample have a stellar mass $M_*$ in the range $10^9 < M_*/M_\odot < 10^{11}$.

More than 100 galaxies have been analysed with the \citetalias{Tremaine1984} method so far, yielding their bar pattern speeds from stellar absorption spectra based on both long-slit and IFU data \citep{Cuomo2020, Williams2021}. In this work, we use the sample provided by \citet{Cuomo2020}, which consists of 104 galaxies since we retain the formally ultrafast bars. Some properties of this sample are shown in Table \ref{Obs_summary_statistics}. Due to the large spread in values, we work in logarithmic space, and also impose limits to the maximum allowed error on the bar length and $\mathcal{R}$ parameter similarly to our main statistical analysis (Section \ref{R_parameter}). This leaves us with 42 galaxies. For ease of comparison, the logarithm of the mean value is then exponentiated, yielding e.g. a typical bar length in kpc rather than $\log$ kpc. The intrinsic dispersion is estimated from a simple population standard deviation neglecting measurement errors, which is justified because the intrinsic dispersion exceeds the root mean square (rms) of the individual fractional measurement errors. However, this is only just true for the crucially important $\mathcal{R}$ parameter, so both measurement errors and intrinsic dispersion need to be simultaneously considered $-$ this is done in Section \ref{R_parameter}.

\begin{table}
	\centering
	\begin{tabular}{lccc}
		\hline 
		Observed & Logarithmic & Intrinsic & rms error \\
		quantity & mean & dispersion (dex) & (dex) \\
		\hline
		$R_{\text{bar}}$ (kpc) & 5.23 & 0.20 & 0.08 \\
		$\mathcal{R}$ & 0.92 & 0.20 & 0.16 \\
		\hline
	\end{tabular}
	\caption{Properties of our adopted observational sample of galaxy bars \citep{Cuomo2020}. We use the (possibly distinct) high and low error bars to get a single logarithmic uncertainty for each quantity for each galaxy, and then apply the same quality cuts as in Section \ref{R_parameter}. This reduces the sample size from 104 to 42. The calculations are done in log-space, and the mean logarithm is then exponentiated to get what we call the logarithmic mean. The intrinsic dispersion is calculated neglecting measurement errors, the rms value of which is shown in the last column. The linear mean value of $\mathcal{R}$ for the 104 (42) galaxies is 1.17 (1.02). Of these galaxies, $\approx 51\% ~(60\%)$ have a reported $\mathcal{R} < 1$.}
	\label{Obs_summary_statistics}
\end{table}

After discarding measurements of $\Omega_p$ with large errors, \citet{Cuomo2020} reported in their section 5.1 that all bars analysed so far with the \citetalias{Tremaine1984} method are compatible with being fast or ultrafast at the 95\% confidence level. The apparently ultrafast bars (responsible for a logarithmic mean $\mathcal{R} < 1$ in Table \ref{Obs_summary_statistics}) are likely due to measurement errors and/or bar-spiral arm alignment leading to an overestimated $R_{\text{bar}}$, as discussed in \citet{Hilmi2020} and \citet{Roshan_2021} in the context of Newtonian and extended gravity theories, respectively. While a slight underestimation of $\mathcal{R}$ is possible in some cases, the bar must be quite fast intrinsically if it appears ultrafast in a careful analysis. Moreover, we explain in detail in the following sections that the employed methods to find the $\mathcal{R}$ parameter in simulated galaxies are identical to those used in observations, so the comparison is in relative terms rather than the true values of $\mathcal{R}$.

\citet{Cuomo2020} suggested that the observed lack of slow bars could be explained if the sample of galaxies analysed so far with the \citetalias{Tremaine1984} method does not include either dynamically evolved barred galaxies or cases with very efficient exchange of angular momentum between the bar and other galaxy components. Since only a small fraction of galaxies are interacting, the more likely scenario seems to be inefficient angular momentum exchange between the bar and any CDM halo. In this work, we explore the feasibility of this in the $\Lambda$CDM model, i.e. whether it necessarily implies efficient slowdown of the bar.

\section{Cosmological hydrodynamical simulations}
\label{cosim}

Using state-of-the-art cosmological hydrodynamical simulation runs of the EAGLE \citep{Schaye_2015, McAlpine_2016} and IllustrisTNG \citep{Nelson_2018, Nelson_2019, Pillepich_2019} projects, we investigate the properties of barred galaxies at $z = 0$ in the $\Lambda$CDM framework. In the following, we briefly describe the here-assessed simulations, and the selection criteria for our galaxy samples. We then describe the method used to estimate the bar strength, length, and corotation radius in simulated galaxies. We show the main steps in our analysis for one illustrative example.

\subsection{The EAGLE simulation}
\label{eagle}

The Evolution and Assembly of GaLaxies and their Environments (EAGLE) project \citep{Schaye_2015, McAlpine_2016} is a set of cosmological simulation runs consistent with \citet{Planck_2014_EAGLE}, i.e. with $H_{0} = 67.77 \, \rm{km\,s^{-1}\,Mpc^{-1}}$, $\Omega_{\mathrm{b},0} = 0.04825$, $\Omega_{\mathrm{m},0} = 0.307$, $\Omega_{\mathrm{\Lambda},0} = 0.693$, $\sigma_{8} = 0.8288$, and $n_{\mathrm{s}} = 0.9611$. The EAGLE simulations were performed with a modification of the \textsc{gadget-3} \citep{Springel_2005} Smoothed Particle Hydrodynamics (SPH) code.

We use the simulations EAGLE Ref-L0050N0752 (hereafter EAGLE50) and EAGLE Ref-L0100N1504 (hereafter EAGLE100), which evolve $2 \times 752^{3}$ (dark matter and baryon) particles in a box with $L = 50$ co-moving Mpc (cMpc) per side and $2 \times 1504^{3}$ particles in a box with $L = 100 \, \rm{cMpc}$ per side, respectively. Both simulations have an initial baryonic particle mass of $1.81 \times 10^{6} \, M_\odot$, a dark matter particle mass of $9.70 \times 10^{6} \, M_\odot$, a co-moving Plummer-equivalent gravitational softening length of $\epsilon_{\mathrm{com}} = 2.66\,\rm{ckpc}$, and a maximum physical softening length of $\epsilon_{\mathrm{phys}} = 0.70 \, \rm{kpc}$. An overview of numerical parameters of different EAGLE simulations is given in tables~2 and 3 of \citet{Schaye_2015}. The EAGLE public database, which includes the EAGLE galaxy database \citep{McAlpine_2016} and EAGLE particle data \citep{TheEAGLETeam_2017}, can be accessed via their website.\footnote{\url{http://icc.dur.ac.uk/Eagle/database.php}}

\subsection{The IllustrisTNG simulation}
\label{TNG}

The Illustris The Next Generation \citep[IllustrisTNG;][]{Marinacci_2018, Naiman_2018, Nelson_2018, Pillepich_2018, Pillepich2_2018, Springel_2018, Nelson_2019, Nelson2_2019, Pillepich_2019} project is a set of cosmological hydrodynamical simulations that further develop the original Illustris project \citep{Genel_2014, Vogelsberger_2014a, Vogelsberger_2014b, Sijacki_2015}. IllustrisTNG consists of various simulation runs with different resolution settings and box sizes. All simulations are consistent with the Planck cosmology \citep{Planck_2016_IllustrisTNG} because they adopt $H_{0} = 67.74 \, \rm{km\,s^{-1}\, Mpc^{-1}}$, $\Omega_{\mathrm{b},0} = 0.0486$, $\Omega_{\mathrm{m},0} = 0.3089$, $\Omega_{\mathrm{\Lambda},0} = 0.6911$, $\sigma_{8} = 0.8159$, and $n_{\mathrm{s}} = 0.9667$. The simulations were performed with the \textsc{arepo} code \citep{Springel_2010}.

Here, we employ the simulations \underline{TNG100-1} and the high-resolution realization TNG50-1, with the -1 suffix indicating the highest available resolution for the box size, and e.g. 100 indicating a box size close to 100 cMpc. The TNG100-1 simulation includes $2 \times 1820^{3}$ initial gas cells and dark matter particles in a co-moving box with length of $L = 75 \, h^{-1} \, \rm{cMpc} \approx 110.7 \, \rm{cMpc}$ per side, where $h$ is the present Hubble constant $H_0$ in units of $100 \, \rm{km\,s^{-1}\, Mpc^{-1}}$. TNG100-1 has a baryonic mass resolution of $m_{\mathrm{baryon}} = 1.4 \times 10^{6} \, M_\odot$, a dark matter mass resolution of $m_{\mathrm{DM}} = 7.5 \times 10^{6} \, M_\odot$, and a Plummer-equivalent gravitational softening length for both the dark matter and stars of $\epsilon_{\mathrm{DM, \, stars}}^{z=0} = 738 \, \rm{pc}$ at redshift $z = 0$. An overview of the TNG100-1 numerical parameters is provided in table~1 of \citet{Nelson_2018}.

The \underline{TNG50-1} simulation has the highest resolution among the here-assessed cosmological simulations, with a resolution of $m_{\mathrm{baryon}} = 8.5 \times 10^{4} M_\odot$, $m_{\mathrm{DM}} = 4.5 \times 10^{5} M_\odot$, and $\epsilon_{\mathrm{DM, \, stars}}^{z=0} = 288 \, \rm{pc}$ \citep[see also table~1 in][]{Pillepich_2019}. It evolves $2 \times 2160^{3}$ initial gas cells and dark matter particles in a box with $ L = 35 \, h^{-1} \, \rm{cMpc} \approx 51.7 \, \rm{cMpc}$ per side. Since we only consider the highest resolution simulation for each TNG box size, we will usually drop the -1 suffix in what follows. The IllustrisTNG data can be downloaded from the dedicated website.\footnote{\url{https://www.tng-project.org/data/}}

In the TNG simulations, an extra flag is defined for each subhalo. The value of this subhalo flag ($subflag$) determines if the galaxy has a cosmological origin ($subflag = 1$), or if it might be a tidal dwarf or CDM-deficient galaxy ($subflag = 0$). Only $\approx 0.13 \%$ of the TNG100 galaxies we use in the likelihood analysis (Section \ref{R_parameter}) are of the second type, a proportion which drops to zero in TNG50 (Table \ref{Sample_sizes}). Therefore, our results are robust with respect to whether we consider only galaxies with a primordial origin. However, we do not apply this restriction in our main analysis because a similar cut is very difficult to apply observationally.

\subsection{Selecting the simulated galaxy sample}
\label{sample-selection}

In all runs, we focus our analysis on galaxies at the $z=0$ snapshot. Our sample includes only galaxies with stellar mass $M_*> 10^{10.0}\, M_\odot$ to ensure that every galaxy has enough resolution to resolve inner structures.

First of all, the discs need to be identified. We find the direction of the total angular momentum vector for particles with $r\leq 0.5 \,r_{1/2}^*$, where $r_{1/2}^*$ is the stellar half-mass radius. We set the $z$-axis along this direction, which we take to be representative of the rotation axis for the inner parts of the galaxy. We then implement two widely used criteria to select disc galaxies:
\begin{enumerate}
    \item $k_{\rm{rot}}\geq 0.5$, where $k_{\rm{rot}}$ is the fraction of stellar kinetic energy in ordered rotation \citep{Sales2010}. This is defined as the mass-weighted average value of $v_{\phi}^2/v^2$ within $30\,$kpc, where $v$ is the total velocity and $v_{\phi}$ is the azimuthal velocity around the rotation axis for each stellar particle. 
    \item The morphological flatness criterion $F\leq 0.7$, where $F \equiv M_1/\sqrt{M_2\, M_3}$ is based on the eigenvalues of the moment of inertia tensor sorted in ascending order so $M_1 \leq M_2 \leq M_3$ \citep{Genel2015}.
\end{enumerate}
The number of discs chosen in this way for different simulations is given in Table \ref{Sample_sizes}. As a consistency check, in the case of TNG100 we apply similar selection rules to those in \citet{Zhao2020}, and we find a similar number of discs.

\begin{table}
	\centering
	\begin{tabular}{ccccc}
		\hline
		\multirow{2}{*}{Simulation} & EAGLE & EAGLE & TNG & TNG \\
		    & 100 & 50 & 100 & 50 \\
		\hline
		Total & 3638 & 481 & 6507 & 903 \\
		Discs & 1512 & 96 & 3586 & 612 \\
		Barred & 141 & 12 & 1166 & 259 \\
		Strong & 19 & 4 & 455 & 126 \\
		Weak & 122 & 8 & 711 & 133 \\
		With reliable $\Omega_p$ & 79 & 6 & 902 & 227 \\
		Used in likelihood & \multirow{2}{*}{70} & \multirow{2}{*}{5} & \multirow{2}{*}{745} & \multirow{2}{*}{209} \\
		 analysis (Section \ref{R_parameter}) & & & & \\
		\hline
	\end{tabular}
	\caption{Bar statistics in different simulations at $z=0$ based on galaxies with stellar mass $M_*> 10^{10.0} \, M_\odot$. The selection rules for identifying discs are $k_{\rm{rot}}\geq 0.5$ and $F \leq 0.7$ (Section \ref{sample-selection}). For consistency, we use the same selection rules for all four simulations considered in this study. The value of $A_2^{\text{max}}$ (Section \ref{bar-st}) is used to classify galaxies as unbarred ($A_2^{\text{max}}<0.2$), weakly barred ($0.2\leq A_2^{\text{max}} < 0.4$), or strongly barred ($A_2^{\text{max}} \geq 0.4$). For TNG100, we recover the statistics presented in \citet{Zhao2020} if we instead require $F\leq 0.5$. In the case of EAGLE100, we find the same number of discs as \citet{Algorry_2017} if we limit ourselves to galaxies with $10^{10.6} \leq M_*/M_\odot \leq 10^{11.0}$ and then require ($k_{\rm{rot}} \geq 0.47, \, F\leq 0.7$) to select discs. However, we find that 30\% of the discs are barred, 31\% of the bars being strong and the other 69\% weak. In contrast, \citet{Algorry_2017} report in their last section that 40\% of the discs are barred, 20\% of these being strong bars and ``another 20\% weak bars''. We presume that the second 20\% is a typo and the authors meant 80\%, in which case there are only small differences in the fraction of barred galaxies and the proportion of these that have weak bars. Of the TNG100 galaxies used in the likelihood analysis, only 1 (ID 469502) out of 745 ($\approx 0.13 \%$) has $subflag = 0$ (for more details, see Section \ref{TNG}). This percentage is zero in TNG50.}
	\label{Sample_sizes}
\end{table}

Among all the disc galaxies, we need to identify which ones host a bar. This means we need to measure the bar strength, which quantifies the non-axisymmetric forces due to the bar potential. To find the $\mathcal{R}$ parameter, we also need to measure the bar length and pattern speed, the latter entering into the corotation radius. In the following, we describe the methods adopted to extract these quantities.

\subsection{Bar strength measurement}
\label{bar-st}

The strength of a bar can be measured as the maximum amplitude of the $\rm{m} = 2$ azimuthal Fourier component used to describe the galaxy surface density \citep{Guo2019}. We divide the disc into annuli of fixed width $\delta r$ and select only disc particles, i.e. those with $\left| z \right| < 1$ kpc. We use $\delta r = 0.1$, 0.2, 0.3, and 0.5 kpc for TNG50, TNG100, EAGLE50, and EAGLE100, respectively. We then compute the Fourier coefficients
\begin{eqnarray}
	a_{\rm{m}} \left( R \right) &\equiv& \frac{1}{M \left( R \right)} \sum_{k=0}^{N} m_k \cos \left( \rm{m} \phi_k \right), ~ \rm{m} = 1, 2, .. \, , \\
	b_{\rm{m}} \left( R \right) &\equiv& \frac{1}{M \left( R \right)} \sum_{k=0}^{N} m_k \sin \left( \rm{m} \phi_k \right), ~ \rm{m} = 1, 2, .. \, , 
\end{eqnarray}
where $N$ is the number of particles in the annulus with mean cylindrical radius $R$ and total annulus mass $M$ consisting of particles labelled by the index $k$ with mass $m_k$ and azimuthal angle $\phi_k$. This lets us define the Fourier amplitude for mode $\rm{m}$ at radius $R$.
\begin{eqnarray}
    A_{\rm{m}} \left( R \right) ~\equiv~ \sqrt{a_{\rm{m}} \left( R \right)^2 + b_{\rm{m}} \left( R \right)^2} \, .
\end{eqnarray}
The bar strength is defined as 
\begin{eqnarray}
    A_2^{\text{max}} ~\equiv~ \max [A_2(R)] \, .
    \label{A2_max}
\end{eqnarray}
The radius at which the maximum occurs can be used as a lower estimate for the bar length.\footnote{We only use this value in the initial estimation of the slit length and height in the \citetalias{Tremaine1984} method, so it is not directly included in our final bar length evaluations.} We divide bars into two categories: strong bars with $A_2^{\text{max}} \geq 0.4$, and weak bars with $0.2 \leq A_2^{\text{max}} < 0.4$. Discs with $A_2^{\text{max}} < 0.2$ are considered unbarred.

\subsection{Fraction of barred galaxies}

\begin{figure}
	\centering
	\includegraphics[width=0.45\textwidth]{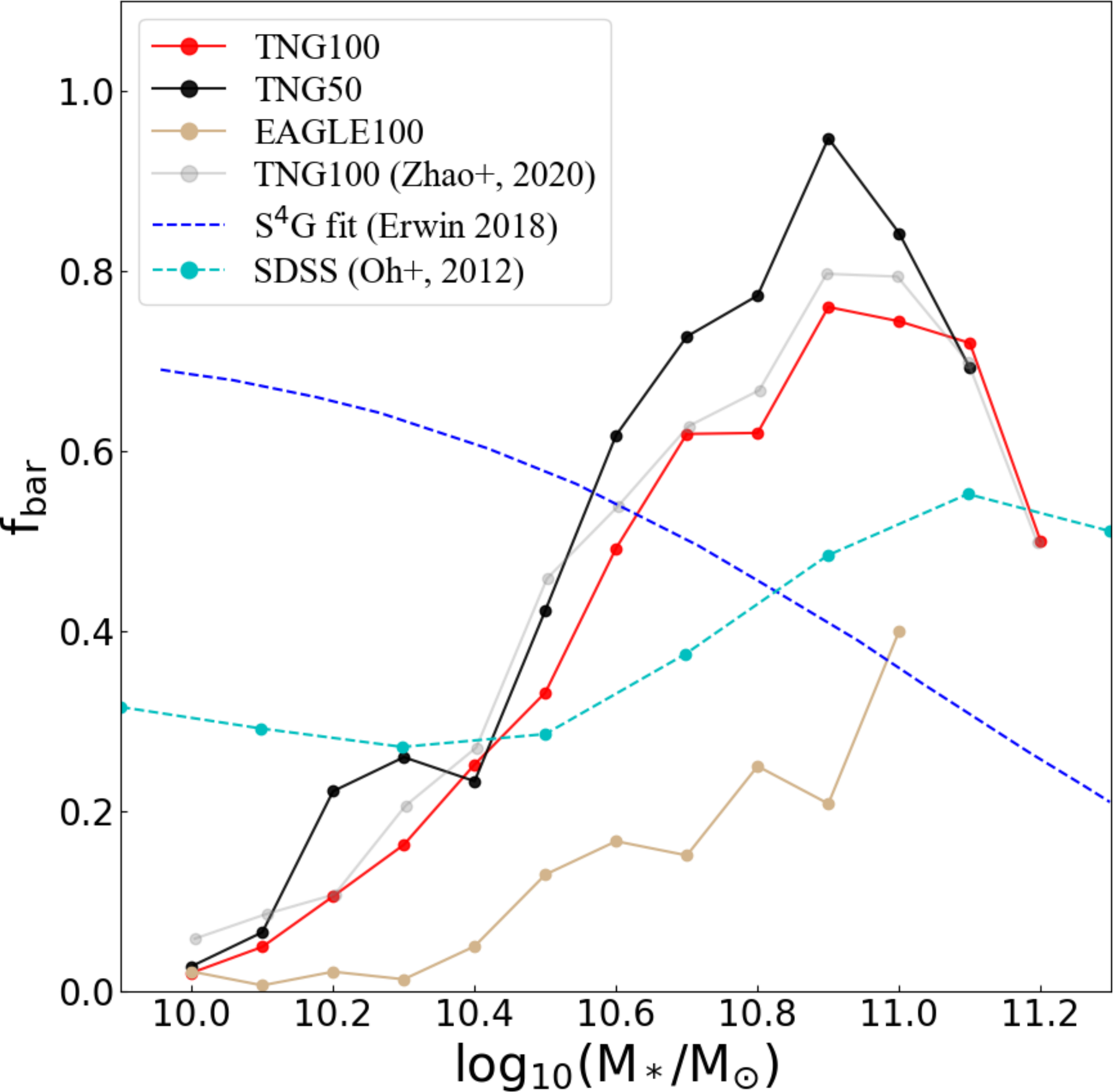}
	\caption{Bar fraction for different simulations. The disc selection rules are $k_{\rm{rot}} \geq 0.5$ and $F \leq 0.5$ (Section \ref{sample-selection}). The red (black) curve shows results for TNG100 (TNG50), while the brown curve shows the bar fraction for EAGLE100. Results for EAGLE50 are not shown due to a very small sample size. The fitted curve to observations (dotted blue) is from \citet{Zhao2020}, while the cyan curve with points shows results from the SDSS \citep{Oh2012}.}
	\label{fig1}
\end{figure}

After requiring $k_{\rm{rot}} \geq 0.5$ and $F \leq 0.5$ to select discs, the bar fraction measured using this approach is shown in Fig. \ref{fig1} for different simulations as a function of stellar mass. We see that compared with TNG100, TNG50 increases the bar fraction almost everywhere in the selected mass range. This might be expected due to the higher resolution in TNG50, but the increase is quite modest and the bar fractions are very similar.

\citet{Erwin2018} highlighted the discrepancy with the bar fraction-stellar mass relation derived from the Spitzer Survey of Stellar Structure in Galaxies (S\textsuperscript{4}G) of the local Universe, and also with most of the SDSS observations that span $z \approx 0.01 - 0.1$. Fig. 5 of their paper shows that in S\textsuperscript{4}G-based observations of galaxies with distances $\la 25$ Mpc, the bar fraction peaks at $M_* \approx 10^{9.7} M_\odot$ and then declines at higher stellar mass (blue curve in Fig. \ref{fig1}). On the other hand, previous SDSS-based studies \citep[e.g.][]{Oh2012} show that the bar fraction is small at $M_* \la 10^{10.5} \, M_\odot$, but continues to grow for larger masses (cyan dashed curve in Fig. \ref{fig1}). \citet{Erwin2018} states that the low fraction of barred galaxies at smaller $M_*$ in SDSS observations is due to low spatial resolution making it difficult to identify bars in lower mass galaxies, a problem which should be greatly alleviated in S\textsuperscript{4}G. 

\citet{Zhao2020} found that in contrast to the bar fraction in S\textsuperscript{4}G observations \citep{Erwin2018}, the fraction of barred galaxies in TNG100 does not follow a similar trend, instead peaking at $M_* \approx 10^{10.9} \, M_\odot$. We show their TNG100 result as the grey curve in Fig. \ref{fig1}. A similar trend is also seen in \cite{Rosas_2020} for the bar fraction in TNG100, and has been argued to have compatibility with some SDSS observations. Furthermore, \citet{Zhou_2020} compared TNG100 with the S\textsuperscript{4}G observational results of \citet{Diaz_2016}, arguing that the bar fraction trend in TNG100 at high stellar mass ($10^{10.66} < M_*/M_\odot < 10^{11.25}$) is in good agreement with S\textsuperscript{4}G observations. However, the discrepancy at lower stellar masses of $M_*/M_\odot < 10^{10.66}$ is also evident in fig. 3 of \citet{Zhou_2020}.

\citet{Zhao2020} show that by using mock SDSS images of TNG100 galaxies, a similar trend to SDSS-based studies is apparent. However, they confirm that bars shorter than 2.5 kpc do not seem to be resolved in SDSS observations, in compliance with the conclusion of \citet{Erwin2018}. Therefore, \citet{Zhao2020} imply that by improving the resolution in the simulations, one might be able to resolve the discrepancy between them and the more reliable bar fractions from S\textsuperscript{4}G observations, especially at the low mass end.

We test this using the black curve in Fig. \ref{fig1}, which shows the bar fraction in the higher resolution simulation TNG50. While the fraction of barred galaxies is increased slightly at almost all stellar masses, the overall picture is hardly changed, so the bar fraction discrepancy reported for TNG100 \citep{Zhao2020} persists into TNG50. This is in line with the recent findings of \citet{Reddish_2021}, highlighting a missing bar problem in the  high-resolution NewHorizon simulation \citep{Dubois_2021} with a resolution of $m_{\mathrm{star}} = 1.3 \times 10^{4} M_\odot$, $m_{\mathrm{DM}} = 1.2 \times 10^{6} M_\odot$, and a spatial resolution of 34~pc. Our findings in TNG50 coupled to these works demonstrate that the problem is not solved either by improving the resolution or by changing the method from moving-mesh to adaptive mesh refinement.

\subsection{Bar length measurement}
\label{bar-ln}

To recover the bar length, we adopt two methods which are commonly used in observational studies: isophotal ellipse fitting and Fourier analysis. The former is based on the radial profile of the ellipticity $\epsilon$ and PA of the ellipses that best fit the galaxy isophotes \citep{Aguerri_2009}. The bar length coincides with the maximum of the ellipticity profile, or to the radius where the PA changes by $\Delta \, \text{PA}=5^\circ$ from the PA of the ellipse with the maximum ellipticity. We select the longer determination in order to infer lower values for the $\mathcal{R}$ parameter, which as we will see later, makes our conclusions more conservative. The isophotal ellipse fitting method is sensitive to the initial conditions, i.e., the semi-major axis and eccentricity used for the first ellipse. Different viable initial conditions are used. The mean value of the length is taken, while the error is their standard deviation.

The other method to find the bar length makes use of the Fourier decomposition of the surface density \citep{Aguerri_2000}. The bar length is computed using the ratio of intensity in the bar ($I_b$) and inter-bar ($I_{ib}$) zones, namely
\begin{equation}
	\mathcal{I} \left( R \right) ~\equiv~ \frac{I_b \left( R \right)}{I_{ib} \left( R \right)} ~=~ \frac{A_0 ~+~ A_2 ~+~ A_4 ~+~ A_6}{A_0 ~-~ A_2 ~+~ A_4 ~-~ A_6 } \, .
	\label{Bar_interbar_ratio}
\end{equation}
The bar length is defined as the outer radius beyond which $\mathcal{I} \left( R \right) < \left( \mathcal{I}^{\text{max}} + \mathcal{I}^{\text{min}} \right)/2$. The error in this case is given by the annular width $\delta r$.

We take the mean of the above two bar length determinations. It is necessary to stress that we implement methods that give upper limits for the bar length. Using other methods (e.g. just using the position of the maximum in $A_2 \left( R \right)$) gives a shorter estimate for the bar length, which would further increase the tension that we identify.

\subsection{Pattern speeds in cosmological simulations}
\label{TW-method}

The bar pattern speed is easily found if viewing a movie of a simulation. However, we need to measure the pattern speed using techniques constructed for real galaxies, where we deal with a single snapshot. The \citetalias{Tremaine1984} method is one of the most precise and widely used methods that allows a direct recovery of $\Omega_p$ when the positions and LOS velocities of the particles/stars are known. The pattern speed is obtained as $\Omega_p \sin i = \langle V \rangle/\langle X \rangle$, where $i$ is the inclination of the disc plane to the sky plane, and 
\begin{eqnarray}
    \langle V \rangle ~&\equiv&~ \frac{\int V_{\text{LOS}} \Sigma \, dX}{\int \Sigma \, dX} \, , \\
    \langle X \rangle ~&\equiv&~ \frac{\int X \Sigma \, dX}{\int \Sigma \, dX}
\end{eqnarray}
are the so-called kinematic and photometric integrals, defined as the luminosity-weighted average LOS velocity $V_{\text{LOS}}$ and position $X$ parallel to the major axis of the particles/stars, respectively, while $\Sigma$ is the corresponding surface density of the tracer. The integrals have to be measured along apertures (or pseudo-slits) parallel to the observed disc major axis projected on the sky.

We find these integrals by using the positions and velocities of the particles in each pseudo-slit to reconstruct the density and LOS velocity maps, thereby mimicking the procedure applied to real data. To measure non-zero integrals, the galaxy should have an intermediate inclination with a bar elongated at an intermediate PA between the projected disc major and minor axes \citep{Cuomo2019b}.

To apply this method, we start with the face-on view of the galaxy, in which we find the bar's PA. We then consider an observer who views the disc at an inclination of $i = 45^\circ$ and whose line of nodes between sky and disc planes is inclined by $60^\circ$ to the bar major axis. We vary the number of evenly spaced slits $N_s$, their width $\Delta_s$, height $h_s \geq N_s \Delta_s$, and length $l_s$ over relatively wide yet reasonable ranges until convergence is reached to the best fit for the pattern speed.

In detail, we start with a fixed combination of $N_s$, $\Delta_s$, and $h_s$, the total extent of the analysed region perpendicular to the slit length direction including also the gaps between slits. We then increase the slit length $l_s$ until the pattern speed converges to a unique value. Linear regression on the kinematic and photometric integrals is used to find the slope $\Omega_p$, while the standard error of the regression is used as the error on $\Omega_p$. Reliable pattern speeds are defined as those with a standard error smaller than 20\%. We then repeat the whole process using different sets of $N_s$, $\Delta_s$, and $h_s$. In this way, we find several estimates for $\Omega_p$.

It is extremely useful to plot $\Omega_p$ against $N_s$ for each galaxy. In this way, one can easily see that the pattern speed in most galaxies has a converging behaviour (an example is shown in Fig. \ref{fig3}). Therefore, it helps to ignore $\Omega_p$ values obtained for those $N_s$ where the pattern speed has not yet converged. The minimum and maximum values of $N_s$ used in our analysis are $5$ and $50$, respectively. We take the mean value of the pattern speed obtained for different values of $N_s$, $\Delta_s$, and $h_s$ as our final estimate for $\Omega_p$, while the largest deviation from the mean value is taken as the error. Instead of taking the mean value, one may take the best-fitting $\Omega_p$ that has the smallest regression standard error. It turns out that this choice does not change the final results. However, we do not recommend it because for some galaxies, the best-fitting $\Omega_p$ occurs for some $N_s$ where convergence has not yet been established in the $\Omega_p-N_s$ plane.

\subsection{An illustrative example}
\label{Example_analysis}

\begin{figure*}
	\centering
	\includegraphics[width=0.7\textwidth]{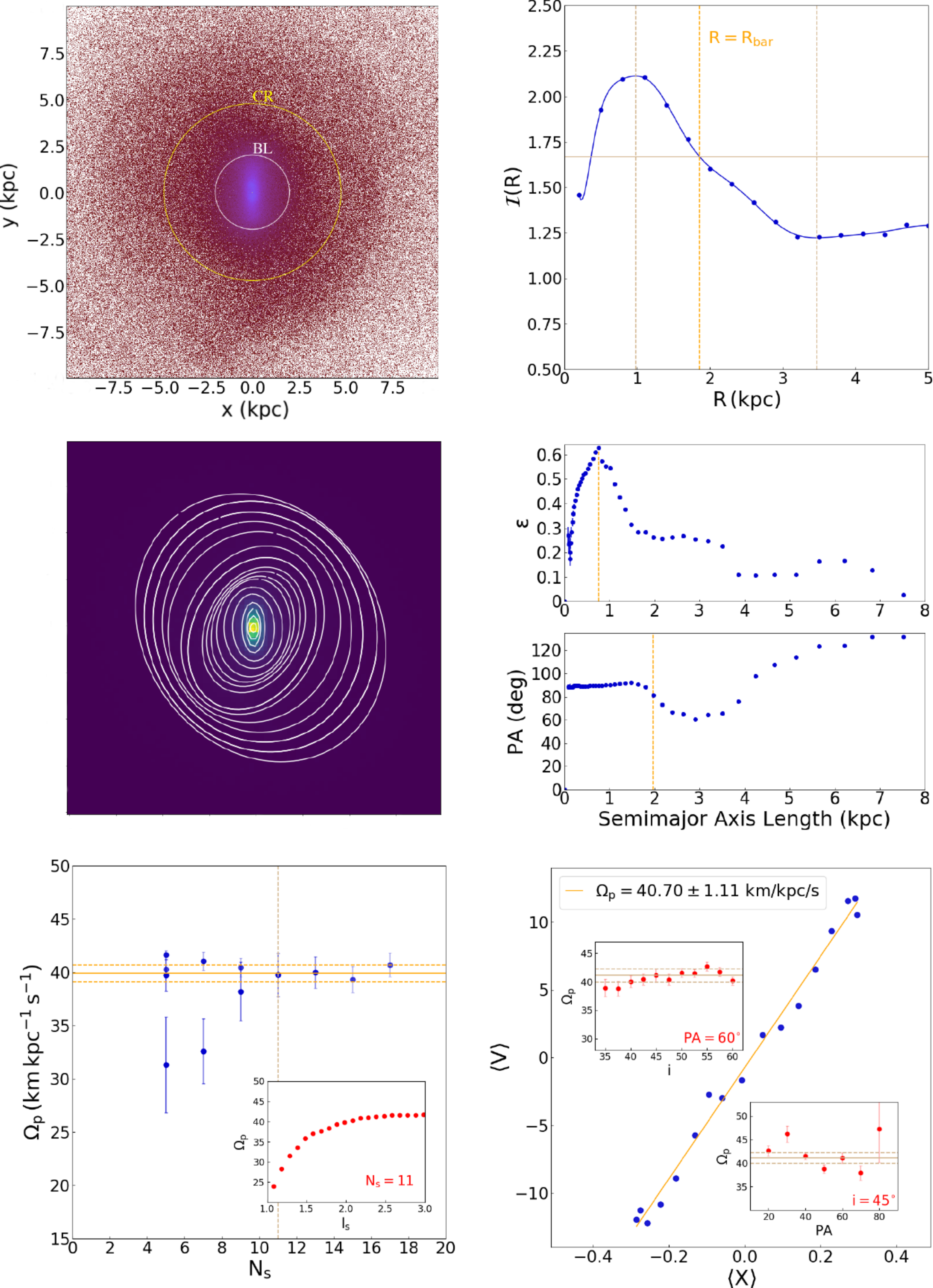}
	\caption{Galaxy ID 229935 in TNG50 as a representative to schematically illustrate the main steps needed to measure $\mathcal{R}$. \emph{Top left}: Face-on view of the stellar particles, with circles showing the bar length (white) and corotation radius (yellow). \emph{Top right}: The intensity ratio between the bar and inter-bar regions (Equation \ref{Bar_interbar_ratio}) around different annuli. \emph{Middle left}: Isophotal ellipse fits. \emph{Middle right}: Ellipticity and position angle of these ellipses as a function of their semi-major axis. \emph{Bottom left}: Inferred $\Omega_p$ as a function of the number of slits $N_s$. The inset fixes $N_s$ at 11 and instead varies the slit length. \emph{Bottom right}: Example of a linear regression between the kinematic and photometric integrals, the slope of which gives $\Omega_p$. The upper and lower insets show, respectively, the reliability of the TW method for different values of the disc inclination $i$ and bar PA relative to the line of nodes between disc and sky planes.}
	\label{fig3}
\end{figure*}

Fig. \ref{fig3} illustrates all the above-mentioned steps to find the $\mathcal{R}$ parameter for a single galaxy in the TNG50 simulation with subhalo ID no. 229935. The face-on projection of the galaxy is demonstrated in the upper left panel, which shows higher density regions using brighter colours. This plot makes use of the \textsc{yt} project \citep{Turk_2011}.

The Fourier method to find the bar length is shown in the upper right panel. The blue curve shows $\mathcal{I}$ as estimated by fitting a 20\textsuperscript{th} order polynomial to the values of ${I_b}/{I_{ib}}$ at several discrete azimuths. The brown vertical dashed lines in this plot show the radii at $\mathcal{I}^{\text{max}}$ and $\mathcal{I}^{\text{min}}$, the maximum and minimum, respectively, of the $\mathcal{I}$ curve. The bar radius is the outer radius at which the horizontal brown line at $(\mathcal{I}^{\text{min}}+\mathcal{I}^{\text{max}})/2$ crosses the $\mathcal{I} \left( R \right)$ curve. This radius is shown by the orange vertical dashed line in the upper right panel. It is our first estimate of the bar length.

The isophotal ellipse fitting method (the second bar length estimation in our analysis) is displayed in the middle panels of Fig. \ref{fig3}. Comparing the upper left and middle left panels, it is apparent that the inner isophotal ellipses are aligned with the bar. Since the bar is located at an angle of $90^{\circ}$, the inner ellipses would also start with a PA of about $90^{\circ}$. This is shown in the lower plot of the middle right panel. In the two plots of this panel, it can be seen that the maximum in the ellipticity curve occurs at radius $R<1$ kpc (upper plot), while the radius at which $\Delta \, \text{PA}= 5^{\circ}$ is at $R\la 2$ kpc (lower plot). Each determination is demonstrated by a vertical orange dashed line in the corresponding panel. As mentioned in Section \ref{bar-ln}, the larger of these two radii would be selected as the bar length in the ellipse fitting method. The final bar length, with which we continue our calculations of the $\mathcal{R}$ parameter, is the mean value of the bar length from the Fourier and the ellipse fitting methods. This average value is shown as the white circle in the upper left panel of Fig. \ref{fig3}.

The lower panels of Fig. \ref{fig3} detail the computation procedure of the \citetalias{Tremaine1984} method in our analysis. As described in Section \ref{TW-method}, to calculate the pattern speed, we select different sets of variables $N_s$, $\Delta_s$, $h_s$, and $l_s$, which are the number of slits, their width, height, and length, respectively. For each set of $N_s$, $\Delta_s$, and $h_s$, we first vary $l_s$ and find the slit length at which $\Omega_p$ converges. Such convergence is demonstrated in the inset to the lower left panel. To check this, we vary the slit length with constant intervals of $0.1$ kpc, and calculate the pattern speed each time. The selected slit length is determined such that the standard deviation of $\Omega_p$ for the ten previous points is less than $10\%$ of the pattern speed's mean value. Furthermore, we recall that to calculate $\Omega_p$ and its error for each case, we use linear regression and its standard error to find the slope of $\langle V \rangle$ as a function of $\langle X \rangle$. An explanatory example is plotted in the lower right panel of Fig. \ref{fig3}. As mentioned in Section \ref{TW-method}, we use intermediate values for the disc inclination and the bar PA, in compliance with previous studies \citep{Debattista_2003, Zou2019}. The reliability of our TW method with respect to different choices for these parameters is examined in the insets to the lower right panel of Fig. \ref{fig3}.

Considering different sets of parameters and choosing only results with a standard error $<20\%$, we plot $\Omega_p$ as a function of $N_s$. This is demonstrated in the lower left panel of Fig. \ref{fig3}, where there exist two converging branches. As explained in Section \ref{TW-test}, since the difference is not large, we only validate the $\Omega_p$ results with $N_s$ larger than the converging point of these branches. This point is shown as the vertical brown dashed line in the lower left panel. The existence of these two branches could be linked to the background features that are visible in the face-on view of the upper left panel. The final $\Omega_p$ and its error are calculated as the mean value of the selected pattern speeds and the largest deviation from the mean, respectively. The final $\Omega_p$ and its error are illustrated as the horizontal solid and dashed lines in the lower left panel.

By combining the final pattern speed with an estimation of the rotation curve as explained in Section \ref{CR}, we find the corotation radius, which is shown as the yellow circle in the upper left panel of Fig. \ref{fig3}.

\subsection{Test of our \citetalias{Tremaine1984} code}
\label{TW-test}

\begin{figure}
	\centering
	\includegraphics[width=0.45\textwidth]{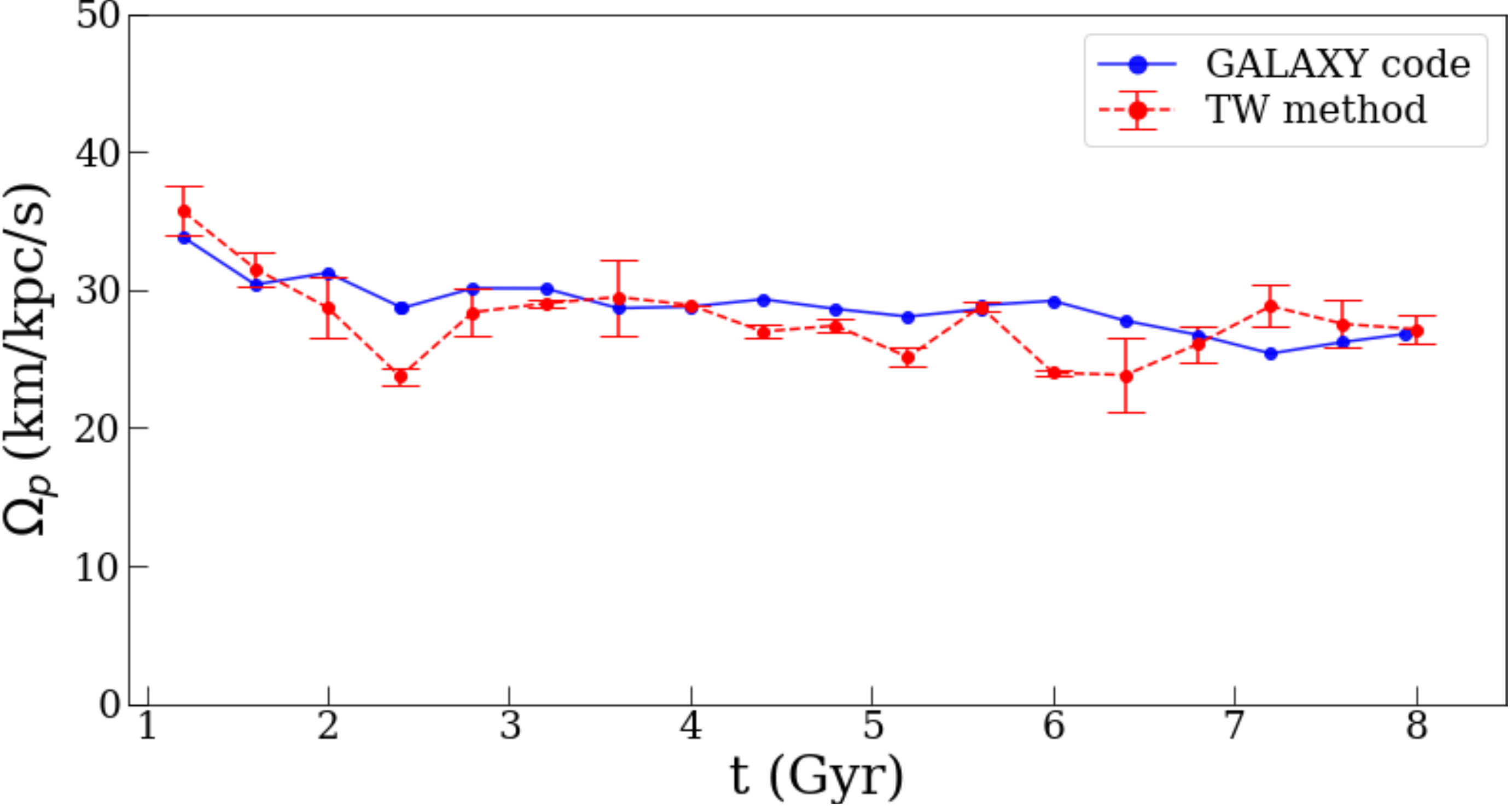}
	\caption{The blue curve indicates the pattern speed for an exponential disc model in the \textsc{galaxy} code \citep{Sellwood_2014}. The red curve is our recovered result using our \citetalias{Tremaine1984} code. The mean fractional error is $<6.7\%$, causing an error on the pattern speed below 1.97 km s\textsuperscript{-1} kpc\textsuperscript{-1}. The deviation is larger when spiral density waves propagate on the disc surface.}
	\label{fig2}
\end{figure}

In order to test if our \citetalias{Tremaine1984} code is able to correctly measure the pattern speed, we simulate an isolated exponential galactic disc with the \textsc{galaxy} code \citep{Sellwood_2014}. In this case, the pattern speed is exactly known as a function of time because results from multiple timesteps are used. In Fig. \ref{fig2}, we compare the correct value of $\Omega_p$ with the one inferred by our \citetalias{Tremaine1984} code at various times. The mean fractional error is $<6.7\%$, which translates to an error on $\Omega_p$ of $\pm 1.97$ km s\textsuperscript{-1} kpc\textsuperscript{-1} $-$ this is a fairly good agreement.

The simulated disc develops spiral features at $t\approx 2.4$ Gyr and $6.4$ Gyr. In these cases, our \citetalias{Tremaine1984} code is not able to correctly identify the pattern speed of the bar, and our estimate deviates by up to 20\% from the true value. A similar deviation due to spiral arms was already pointed out by \citet{Hilmi2020}. Therefore, we carefully treat galaxies that have spiral structures. These galaxies typically yield more than one $\Omega_p$ using the \citetalias{Tremaine1984} method. Such behaviour can also happen in observations \citep{Rand2004}. For these galaxies, we see at least two converging branches in the $\Omega_p-N_s$ plane. When the difference between the branches is large, we categorize the galaxy as not having a reliable pattern speed. Otherwise, the mean value of $\Omega_p$ described in Section \ref{TW-method} gives a reliable pattern speed. In most galaxies with the above-mentioned bifurcation, our approach gives larger errors for $\Omega_p$ compared to galaxies without any spiral structures, so in principle, our analysis automatically inflates the uncertainties when the technique is less reliable.

\subsection{Rotation curve and corotation radius measurement}
\label{CR}

To find the corotation radius $R_{\text{CR}}$ of each galaxy, we need to measure its rotation curve $v_c \left( R \right)$ in order to solve the implicit equation $\Omega_p R = v_c \left( R \right)$, the non-zero solution to which gives the corotation radius. The rotation curve is found from the rms azimuthal velocity $\sqrt{\langle v_{\phi}^2\rangle}$ in terms of $R$ for particles that belong to the disc ($\vert z\vert<1\,$kpc). This gives a first estimation of the galaxy's rotation curve, which we call $v_{\phi}^{\text{rms}} \left( R \right)$. Adding the extra condition that the considered particles should move on nearly circular orbits (with $v_{\phi}^2/v^2\geq 0.9$) gives a second and much better estimation, which we call $v_{c} \left( R \right)$. In this way, we ignore the existence of azimuthal velocity dispersion and the asymmetric drift correction.

Our approach recovers the true rotation curve $v^{*}_c(R)$ fairly well, which we determine in some cases using a much more computationally expensive procedure by calculating the radial gravity produced by all the mass components of the galaxy at a suitable grid of positions in the disc plane. Another way to measure the rotation curve is $\hat{v}_c \left( R \right) \equiv \sqrt{GM \left( R \right)/R}$, where $M \left( R \right)$ is the total (baryonic and dark halo) mass inside radius $R$. Using the estimates $v_{\phi}^{\text{rms}} \left( R \right)$ or $v_{c} \left( R \right)$ has the advantage that only the star particles would be enough, so we do not need data on the CDM and gas particles to measure the rotation curve. In addition to greatly reducing the computational cost, it is satisfactory in the sense that our estimated pattern speeds also use only star particles as tracers for the kinematic and photometric integrals. We therefore choose $v_c$ as the main measure for the rotation curve in this paper.

\begin{figure*}
	\centerline{\includegraphics[width = 5.7cm]{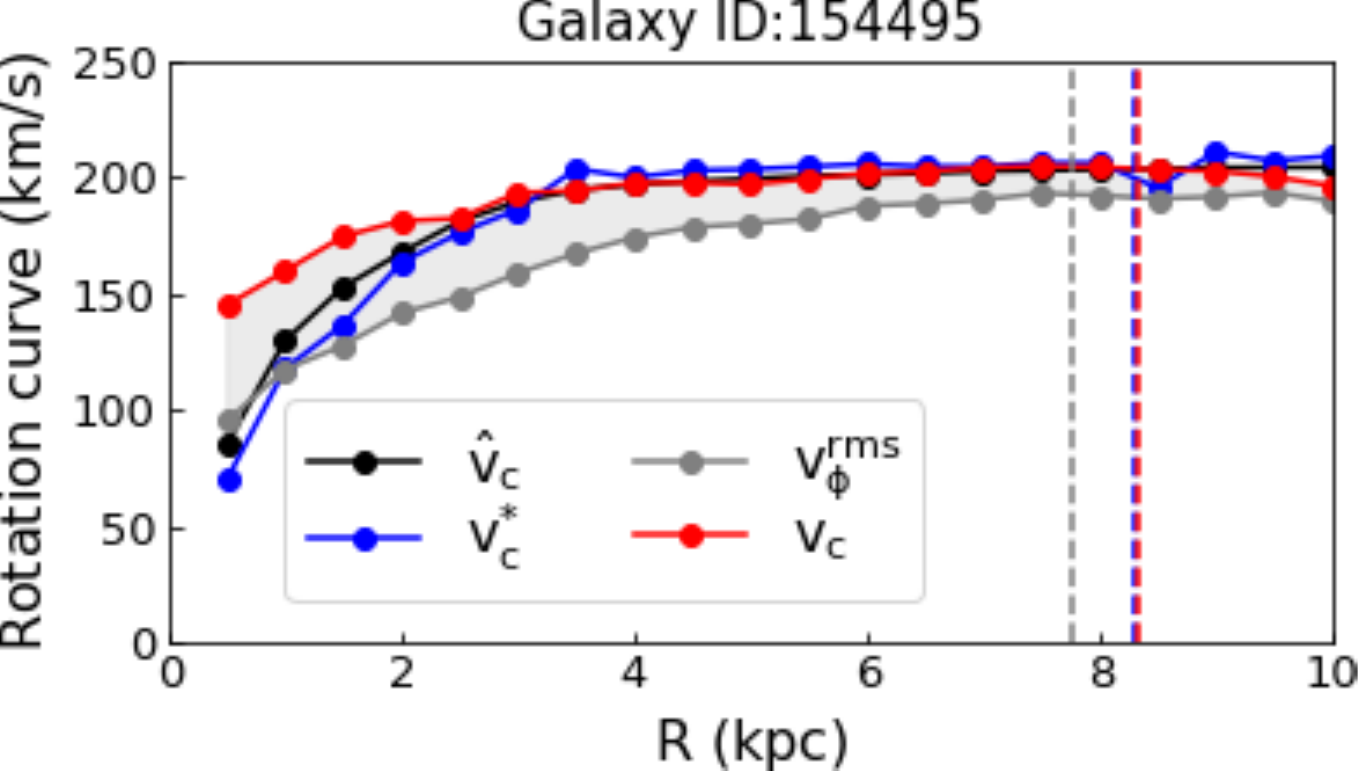}
	\hspace{0.3cm}
	\includegraphics[width=5.7cm]{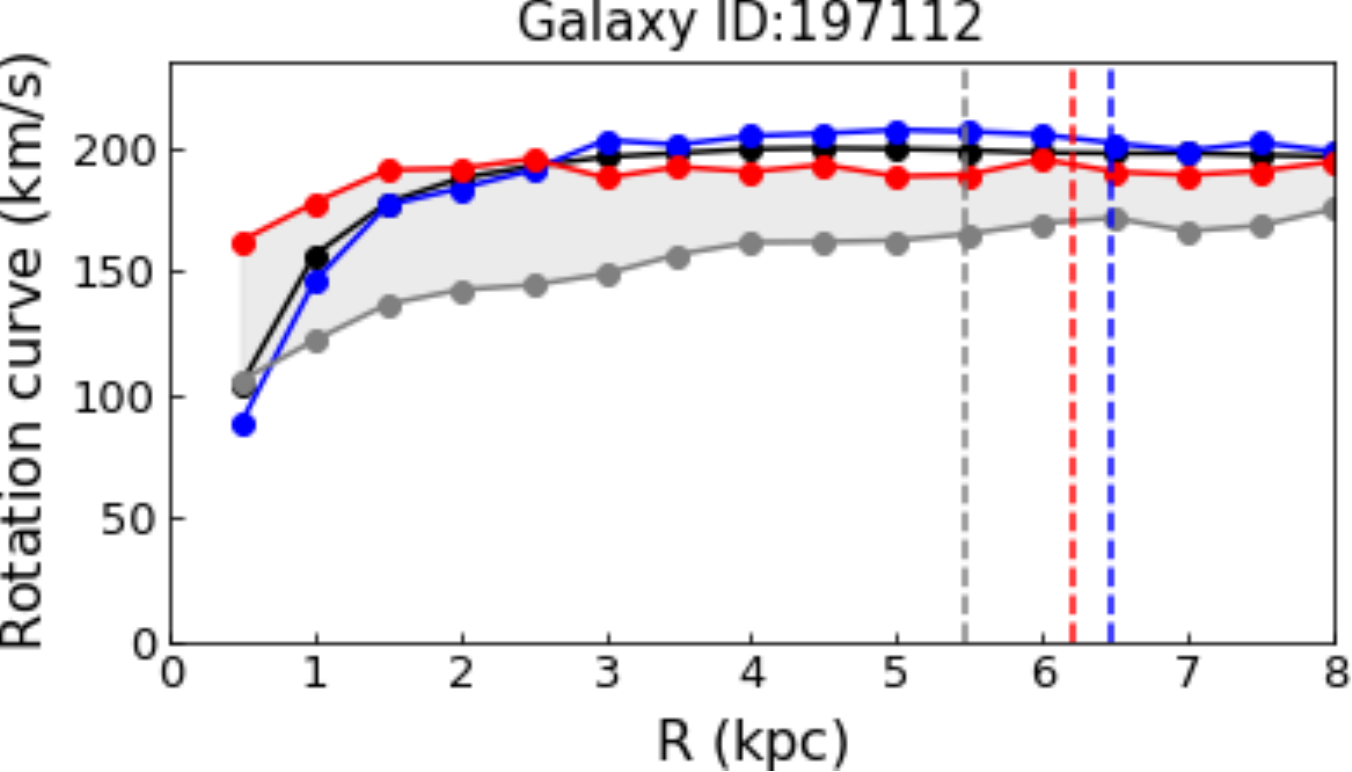}
	\hspace{0.3cm}
	\includegraphics[width=5.7cm]{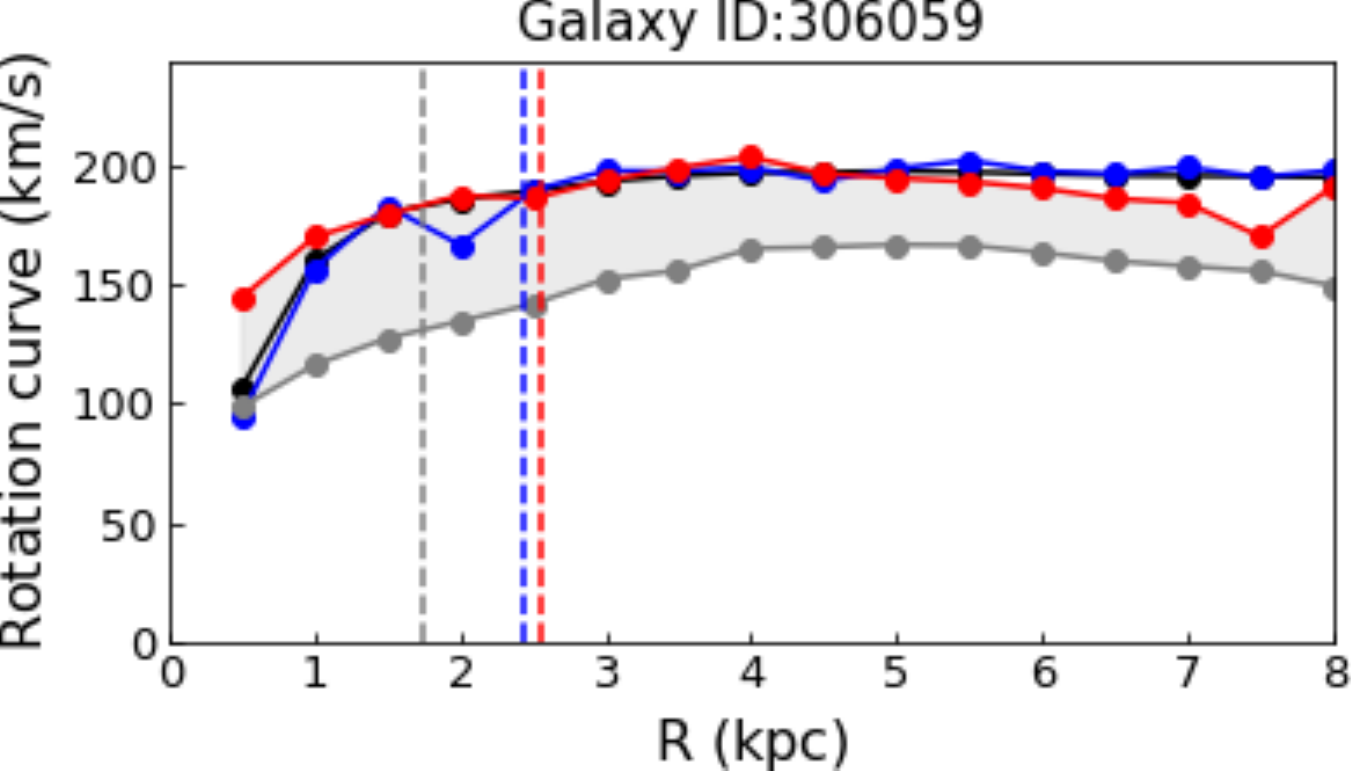}}
	\vspace{0.2cm}
	\centerline{\includegraphics[width = 5.7cm]{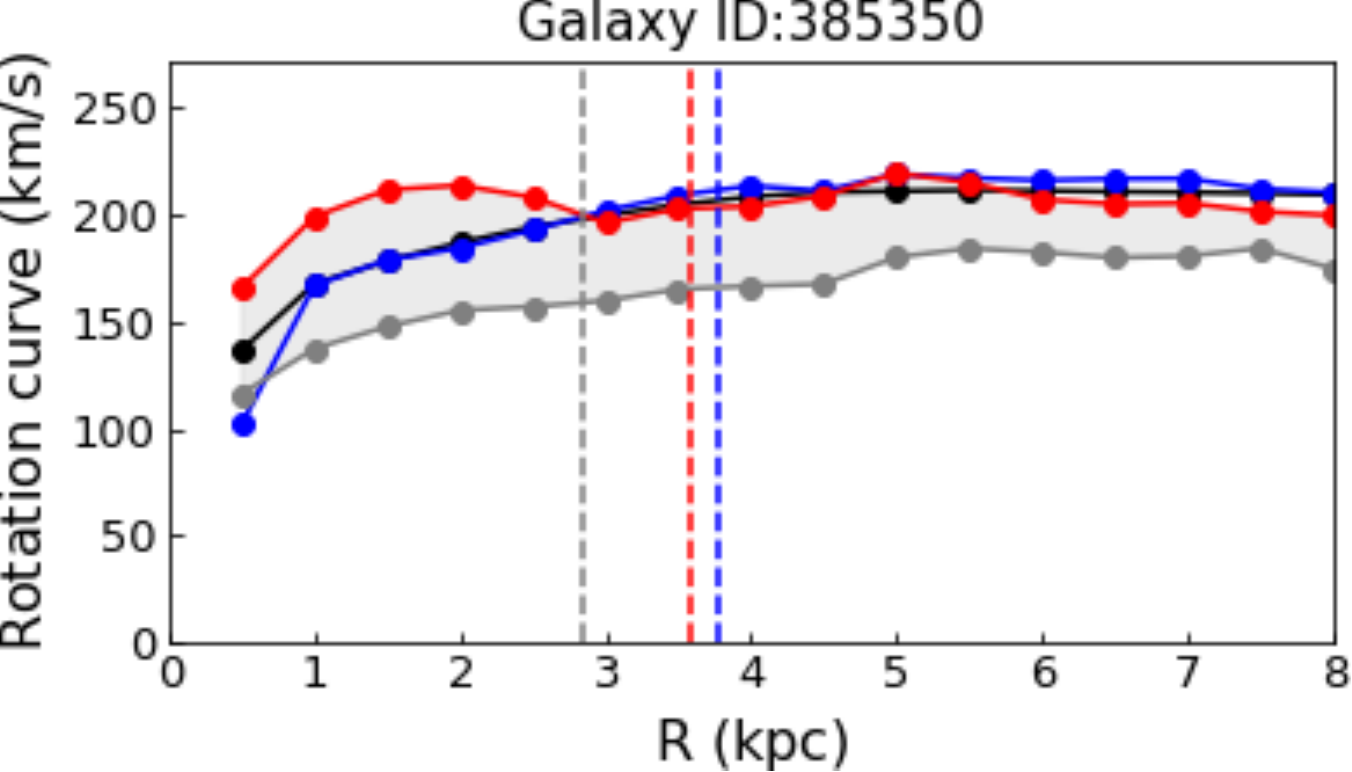}
	\hspace{0.3cm}
	\includegraphics[width=5.7cm]{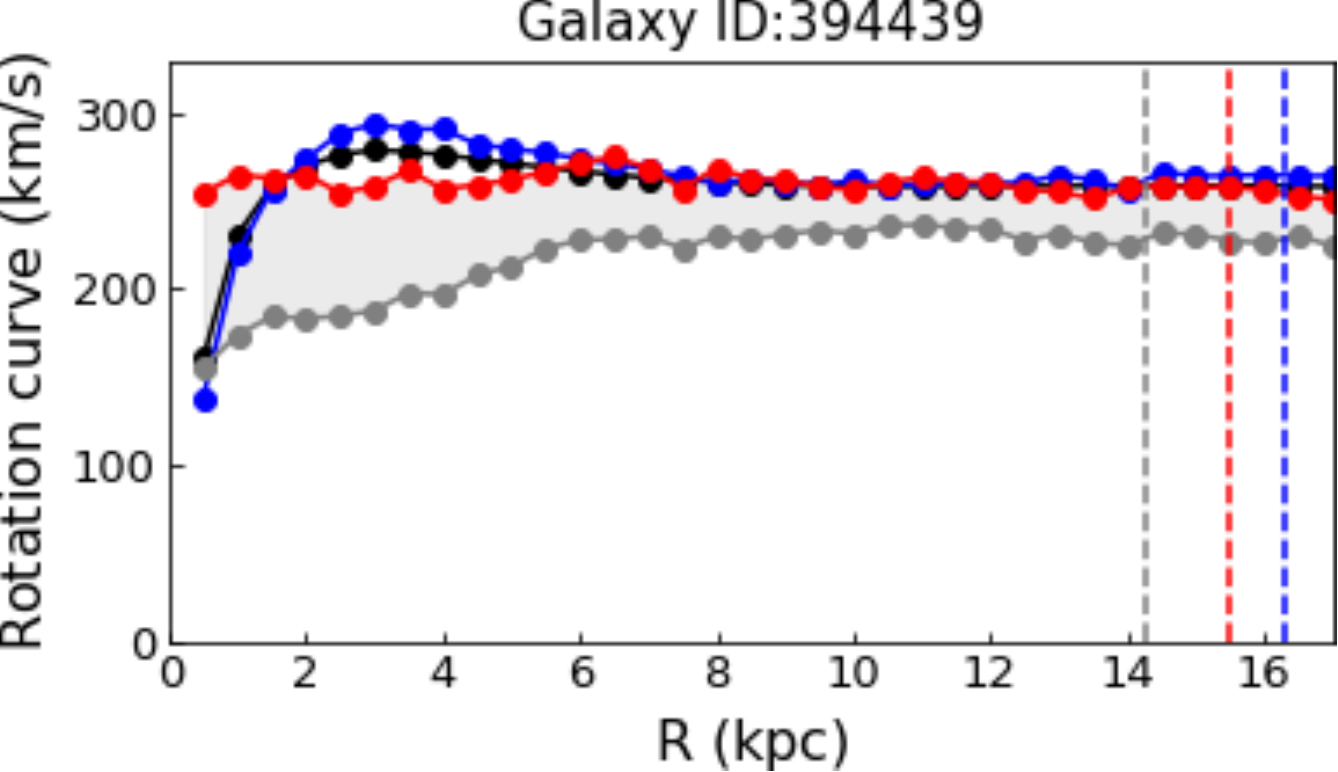}
	\hspace{0.3cm}
	\includegraphics[width=5.7cm]{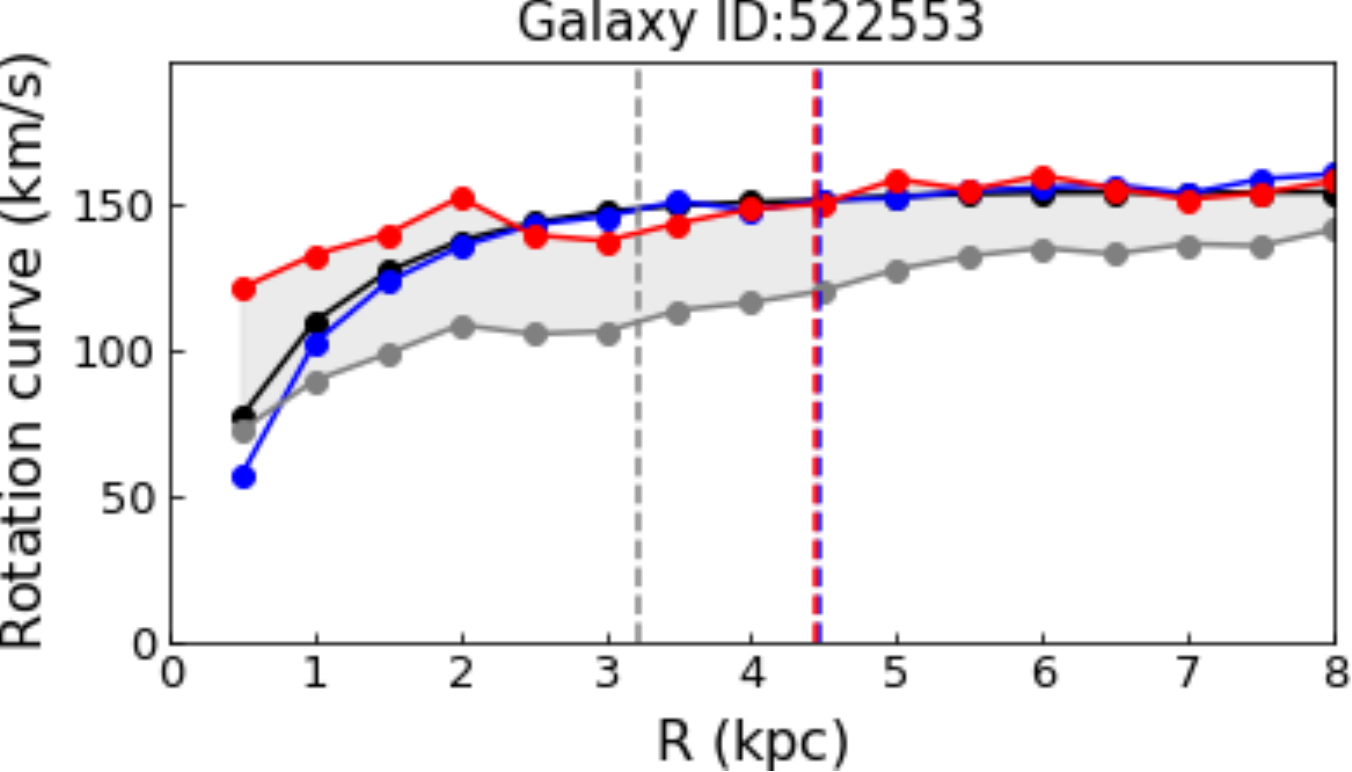}}
	\caption{The rotation curve as calculated using different methods (Section \ref{CR}). We use the red and grey curves as high and low estimates, respectively. The vertical dashed lines are the corresponding corotation radii, with the colour matching that of the used rotation curve. The black and blue rotation curves are based on force calculations that use the total matter distribution, including the CDM particles (see the text). The most accurate method should be $v^{*}_c$ (blue curve), but this is quite expensive to calculate in all cases, and would differ from observational studies that do not directly measure CDM. Since $v^{*}_c$ matches the $v_c$ curve (red) most closely, we use $v_c$ as our nominal choice instead of $v_{\phi}^{\text{rms}}$ (grey curves).}
	\label{RC}
\end{figure*}

To test the accuracy of our choice, we selected three strongly barred and three weakly barred galaxies with a variety of pattern speeds ranging from $\approx 16.3 - 73.6$ km s\textsuperscript{-1} kpc\textsuperscript{-1}. We compared $v_{c}$ and $v_{\phi}^{\text{rms}}$ (both available for our full sample) with $v^{*}_c$ and $\hat{v}_c$. The results are presented in Fig. \ref{RC}. It is clear that $\hat{v}_c$ (the black curves) and $v_c^*$ (the blue curves) match very well throughout the disc. In addition, we see that $v_c$ (the red curves) also recover $\hat{v}_c$ (and $v_c^*$) to an average accuracy of better than 4\% for $R\ga 2\,$kpc. The accuracy is lower ($\approx 15\%$) at $R\la 2\,$kpc. However, only a very small fraction of galaxies have a corotation radius smaller than $2\,$kpc. Therefore, our decision to use $v_c$ should be appropriate for most of the galaxies we study.

For the rare cases where $R_{\text{CR}}$ is small, $v_c$ overestimates the rotation curve. Consequently, the corresponding corotation radius will be overestimated. This increases the inferred $\mathcal{R}$ parameter, which may enhance the tension with observations in a fake way. To completely suppress this possibility, we also measure $R_{\text{CR}}$ using $v_{\phi}^{\text{rms}}$, represented by the grey curves in Fig. \ref{RC}. Notice that $v_{\phi}^{\text{rms}}$ almost always underestimates the rotation curve and thus the corotation radius, which would lead to an underestimated $\mathcal{R}$ parameter that artificially reduces the tension. However, we prove in Section \ref{newCR} that using $v_{\phi}^{\text{rms}}$ does not alter our main conclusion that the observed $\mathcal{R}$ parameter distribution rules out the standard dark matter simulations. This is important because it is clear from Fig. \ref{RC} that the filled area between the red and grey curves reliably brackets the true rotation curve ($v_c^*$) at almost all radii.

For a better illustration of the uncertainties in the corotation radius, we also plot this as derived from our two generally available estimations of the rotation curve ($v_c$ and $v_{\phi}^{\text{rms}}$), and compare them with the estimate using the true rotation curve ($v_c^*$) for each of the six galaxies in Fig. \ref{RC}. Each estimated corotation radius is plotted as a dashed vertical line with the same colour as the rotation curve used to calculate it. Our decision to use $v_c$ is based on the fact that corotation radii measured this way (the red dashed line in each panel) gives a closer estimation to the true value based on $v_c^*$, shown as the blue line. This underlines that $v_c$ should be a more reliable rotation curve measure than $v_{\phi}^{\text{rms}}$ in the outer regions relevant to corotation radius measurements. As a final remark, let us mention that the mean fractional error of the red corotation radii (our adopted values) in comparison to the blue ones ($v_c^*$) is only $\approx 3.41\%$ in these six galaxies.

\section{Results}
\label{res}

\subsection{Bar properties in the IllustrisTNG simulation}

\begin{figure*}
	\centering
	\includegraphics[width=0.49\textwidth]{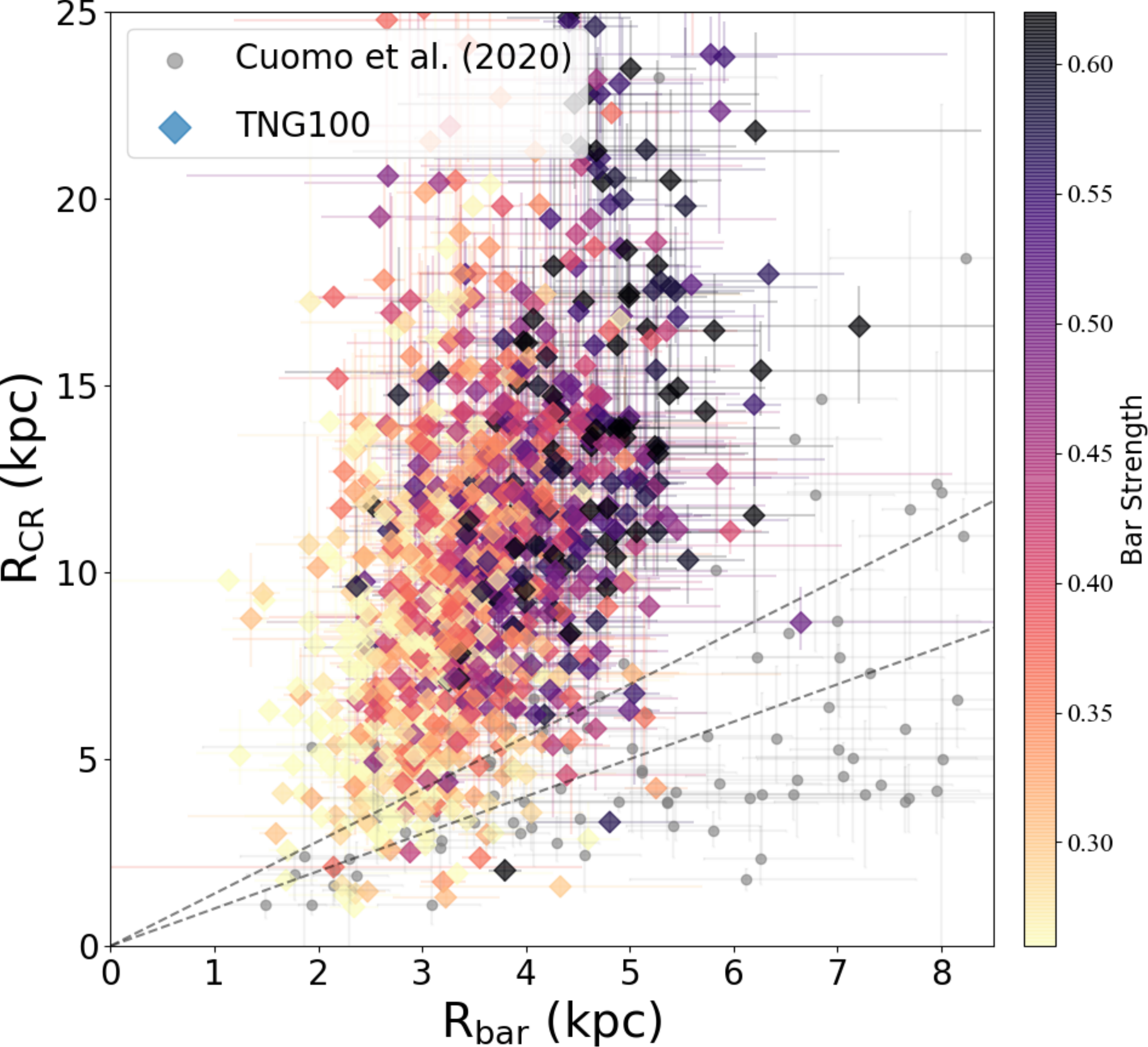}
	\includegraphics[width=0.49\textwidth]{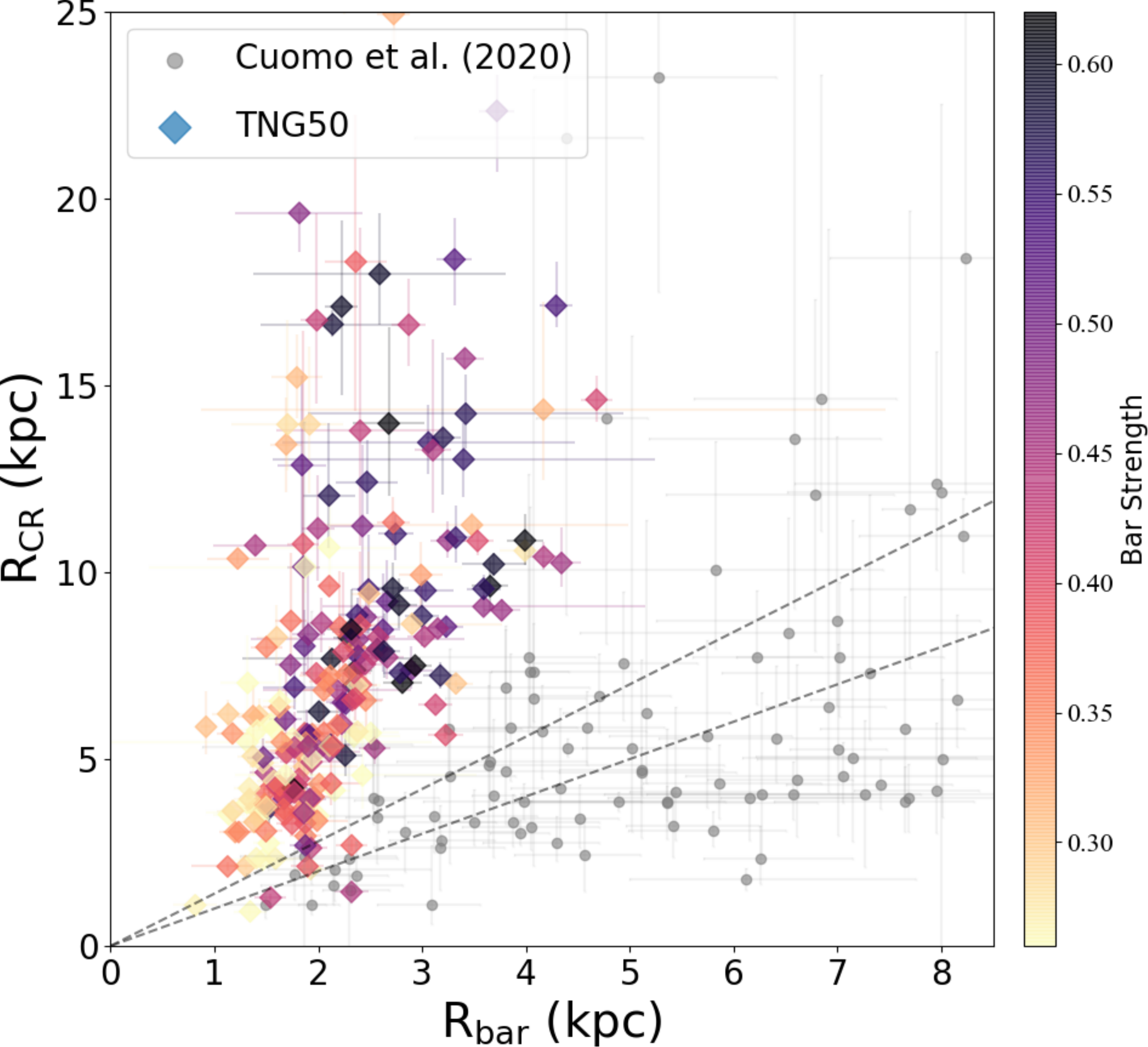}
	\caption{Relation between bar corotation radius and length for IllustrisTNG at $z=0$ (diamond markers). We show results for the barred galaxies in TNG100 (\emph{left}) and TNG50 (\emph{right}). The colour shows the bar strength $A_2^{\text{max}}$ (Section \ref{bar-st}), as shown on the colour bar. The solid grey circles indicate observed data from \citet{Cuomo2020}. The \citetalias{Tremaine1984} method has been used to measure the pattern speeds. The dashed lines indicate the borders of the fast bar regime ($\mathcal{R} = 1-1.4$).}
	\label{fig4}
\end{figure*}

We restrict ourselves to the TNG100 and TNG50 main runs in IllustrisTNG.\footnote{These are denoted with a suffix -1 in the technical literature, because lower resolution simulations with the same box size also exist for comparison purposes $-$ but are not considered here.} In all of our samples, we select galaxies with stellar mass $M_*> 10^{10.0} \, M_\odot$. Using our selection rules (Section \ref{sample-selection}), the number of barred galaxies is shown in Table \ref{Sample_sizes}. We use Fig. \ref{fig4} to plot the bar corotation radius against the bar length for all TNG100 and TNG50 barred galaxies, with the colour of each point indicating the bar strength, i.e. the maximum amplitude over radius of the normalized azimuthal $\rm{m} = 2$ Fourier component of the surface density within each annulus (Equation \ref{A2_max}). The dashed lines indicate the borders of the ultrafast bar ($\mathcal{R}<1$) and slow bar ($\mathcal{R}>1.4$) regimes, with the region between them ($\mathcal{R} = 1 - 1.4$) known as the fast bar regime. The observational results from \citet{Cuomo2020} are shown as grey points. It is clear that unlike real galaxies, most barred TNG galaxies have a slow bar. Moreover, stronger bars tend to be slower compared to weaker bars, which is to be expected because a stronger bar should create a stronger response in the dark halo. A positive correlation between bar length and bar strength is seen in both TNG100 and TNG50, and is consistent with the numerical simulations of \citet{Klypin_2009}.\footnote{For further discussion of previous isolated CDM simulations of galaxy bars, we refer the reader to section 5.4.1 of \citet{Roshan_2021}.}

\begin{table*}
	\centering
	\begin{tabular}{lcccccc}
		\hline
		Simulation & \multicolumn{3}{c}{TNG100} & \multicolumn{3}{c}{TNG50} \\
		\hline 
		& Mean & Intrinsic & rms error & Mean & Intrinsic & rms error \\
		Quantity & value & dispersion (dex) & (dex) & value & dispersion (dex) & (dex) \\
		\hline
		$\Omega_p$ & 19.58 & 0.23 & 0.07 & 30.79 & 0.24 & 0.04 \\
		$R_{\text{bar}}$ & 3.43 & 0.11 & 0.06 & 2.05 & 0.13 & 0.05 \\
		$\mathcal{R}$ & 2.77 & 0.20 & 0.12 & 3.03 & 0.20 & 0.08 \\
		\hline
	\end{tabular}
	\caption{Similar to Table \ref{Obs_summary_statistics}, but for bars in the IllustrisTNG simulations (Section \ref{TNG}). Calculations are done in log-space after imposing quality cuts similarly to Section \ref{R_parameter}. The mean is then exponentiated.}
	\label{TNG_summary_statistics}
\end{table*}

Table \ref{TNG_summary_statistics} shows a few summary statistics for bars in the IllustrisTNG simulation, calculated similarly to Table \ref{Obs_summary_statistics} (i.e. in log-space, but the mean logarithmic value is then exponentiated for convenience). The mean pattern speed in TNG100 (TNG50) is $\overline{\Omega}_p=19.58\,$km s\textsuperscript{-1} kpc\textsuperscript{-1} (30.79~km s\textsuperscript{-1} kpc\textsuperscript{-1}). While bar pattern speeds are higher in TNG50, bars are meaningfully shorter here. In particular, galaxies in TNG100 with a well-defined pattern speed have a mean bar length of $\overline{R}_{\text{bar}}=3.43\,$kpc, while this is only $\overline{R}_{\text{bar}}=2.05\,$kpc in TNG50. Therefore, bars are 40\% shorter in TNG50. This increases the $\mathcal{R}$ parameter, though both TNG50 and TNG100 give a logarithmic mean $\mathcal{R}$ much higher than in observations (Table \ref{Obs_summary_statistics}) $-$ we quantify the tension in Section \ref{R_parameter}. We stress that the higher logarithmic mean $\mathcal{R}$ in TNG50 is already a sign that the bar speed tension is not reduced by its higher resolution compared to TNG100 (Section \ref{TNG}) $-$ the change is in the wrong direction to bring about agreement with observations, where $\mathcal{R}$ is typically close to 1 (Table \ref{Obs_summary_statistics}).

\subsection{Bar properties in the EAGLE simulation}
\label{rp}

\begin{figure}
	\centering
	\includegraphics[width=0.45\textwidth]{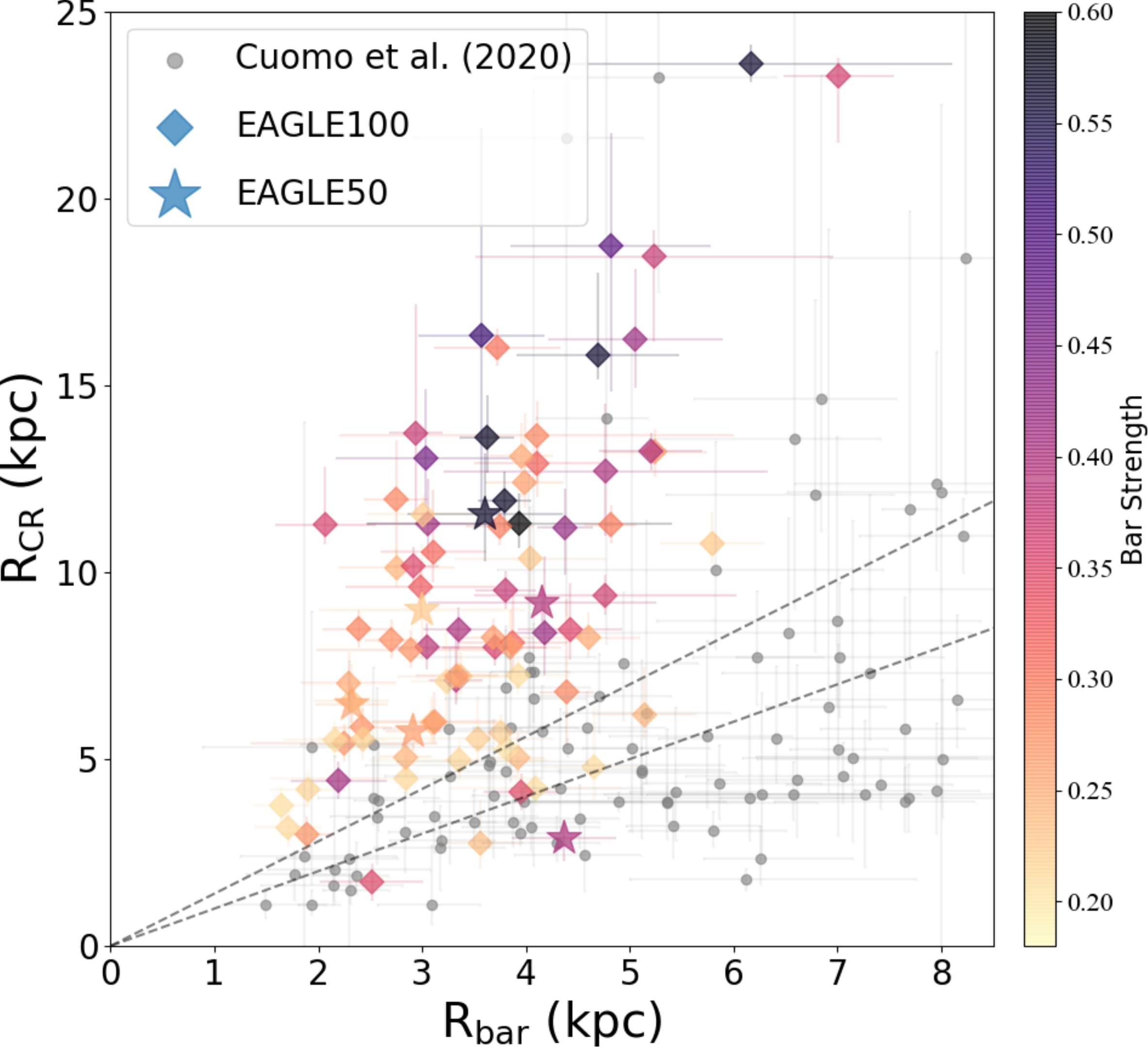}
	\caption{Similar to Fig. \ref{fig4}, but now showing barred galaxies in EAGLE100 (diamond markers) and EAGLE50 (star markers) at $z=0$.}
	\label{fig5}
\end{figure}

In addition to the more recent TNG100 and TNG50, we include the earlier EAGLE simulation in our analysis for the sake of completeness. This is of great importance in helping to check if similar bar statistics are obtained using a somewhat different simulation setup to the IllustrisTNG project, though still within the $\Lambda$CDM paradigm. In the EAGLE case, we focus on the runs EAGLE100 and EAGLE50 because EAGLE25 is too small to permit meaningful statistics. We measure all the quantities using the same methods as for the IllustrisTNG case. The relation between corotation radius and bar length for EAGLE galaxies is shown in Fig. \ref{fig5}. In both runs, most galaxies lie in the slow bar regime. As in the IllustrisTNG case, stronger bars tend to be longer and slower.

Similarly to the IllustrisTNG case, we use Table \ref{EAGLE_summary_statistics} to show a few summary statistics for the bars in EAGLE galaxies that have one. The mean value of the bar length in EAGLE100 and EAGLE50 is $\overline{R}_{\text{bar}}=3.44\,$kpc and $\overline{R}_{\text{bar}}=3.38\,$kpc, respectively, so the difference is small. The bar pattern speed differs somewhat $-$ its mean value is $\overline{\Omega}_p=24.34\,$km s\textsuperscript{-1} kpc\textsuperscript{-1} in EAGLE100 and $\overline{\Omega}_p=32.25\,$km s\textsuperscript{-1} kpc\textsuperscript{-1} in EAGLE50. However, given the small sample size (Table \ref{Sample_sizes}) and the significant intrinsic dispersion, this might just be a random fluctuation. One reason for the lack of difference between EAGLE100 and EAGLE50 could be that they use the same resolution (Section \ref{eagle}), while TNG50 has a much higher resolution than TNG100 (Section \ref{TNG}).

\begin{table*}
	\centering
	\begin{tabular}{lcccccc}
		\hline
		Simulation & \multicolumn{3}{c}{EAGLE100} & \multicolumn{3}{c}{EAGLE50} \\
		\hline 
		& Mean & Intrinsic & rms error & Mean & Intrinsic & rms error \\
		Quantity & value & dispersion (dex) & (dex) & value & dispersion (dex) & (dex) \\
		\hline
		$\Omega_p$ & 24.34 & 0.22 & 0.05 & 32.25 & 0.17 & 0.05 \\
		$R_{\text{bar}}$ & 3.44 & 0.13 & 0.07 & 3.38 & 0.11 & 0.08 \\
		$\mathcal{R}$ & 2.42 & 0.16 & 0.11 & 1.91 & 0.27 & 0.12 \\
		\hline
	\end{tabular}
	\caption{Similar to Table \ref{TNG_summary_statistics}, but for bars in the EAGLE simulations (Section \ref{eagle}).}
	\label{EAGLE_summary_statistics}
\end{table*}

\section{Statistical distribution of the \texorpdfstring{$\mathcal{R}$}{R} parameter}
\label{R_parameter}

The parameter $\mathcal{R}$ (Equation \ref{R_definition}) specifies the `speed' of the bar in dimensionless form, with large values defined as a slow bar. The logarithmic mean value of this parameter for the sample of observed galaxies from \citet{Cuomo2020} is $\overline{\mathcal{R}}=0.92$. In the cosmological simulations, we have $\overline{\mathcal{R}} = 2.77$, $3.03$, $2.42$, and $1.91$ for TNG100, TNG50, EAGLE100, and EAGLE50, respectively. The discrepancy is probably not due to resolution effects because the highest resolution run (TNG50) has the worst compatibility with observations.

Although the discrepancy is clearly evident from the mean values of $\mathcal{R}$, one needs to carefully find its distribution in both observations and simulations in order to quantify the tension between them. To find the posterior inference on $\mathcal{R}$ and its intrinsic dispersion, we follow the prescription presented in section 4.6 of \citet{Roshan_2021}, which we briefly summarize here. The main difference with their work is that they neglected uncertainties in simulated values of $\mathcal{R}$ and thus considered errors only on the observational side, whereas here we consider errors on both simulated and observed $\mathcal{R}$ values. As in \citet{Roshan_2021}, we assume $\mathcal{\widetilde{R}} \equiv \log_{10} \mathcal{R}$ is distributed as a Gaussian with mean $\overline{\mathcal{\widetilde{R}}}$ and intrinsic dispersion $\sigma_{\mathcal{\widetilde{R}}}$. We infer the population parameters $\left(\overline{\mathcal{\widetilde{R}}}, \sigma_{\mathcal{\widetilde{R}}} \right)$ from observations \citep{Cuomo2020} and using different cosmological simulations (Section \ref{cosim}), allowing us to quantify the tension between them.

To calculate the uncertainty on the $\mathcal{R}$ parameter in each simulated or observed galaxy, we average the low and high error bars to come up with a single uncertainty $\delta R_{\text{bar}}$ for the measured bar length and $\delta R_{\text{CR}}$ for its corotation radius. We then require $R_{\text{bar}}$ and $R_{\text{CR}}$ to both have a fractional uncertainty $<\epsilon = \frac{1}{3}$. The analogous criterion is imposed on the corotation radius $R_{\text{CR}}$ and its uncertainty $\delta R_{\text{CR}}$.

We use the value and uncertainty for each corotation radius and bar length to estimate the fractional uncertainty $\alpha$ in their ratio $\mathcal{R}$.
\begin{eqnarray}
	\alpha ~\equiv~ \frac{\delta \mathcal{R}}{\mathcal{R}} ~=~ \sqrt{\left(\frac{\delta R_{\text{bar}}}{R_{\text{bar}}} \right)^2 + \left(\frac{\delta R_{\text{CR}}}{R_{\text{CR}}} \right)^2} \, .
\end{eqnarray}
To further assure the quality of our dataset, we require that $\alpha < \epsilon$ and estimate $\sigma_i$ as
\begin{eqnarray}
	\sigma_i ~=~ \frac{1}{2} \log_{10} \left( \frac{1 + \alpha}{1 - \alpha} \right) \, .
	\label{Averaged_log_error}
\end{eqnarray}

\begin{figure}
	\centering
	\includegraphics[width=0.45\textwidth]{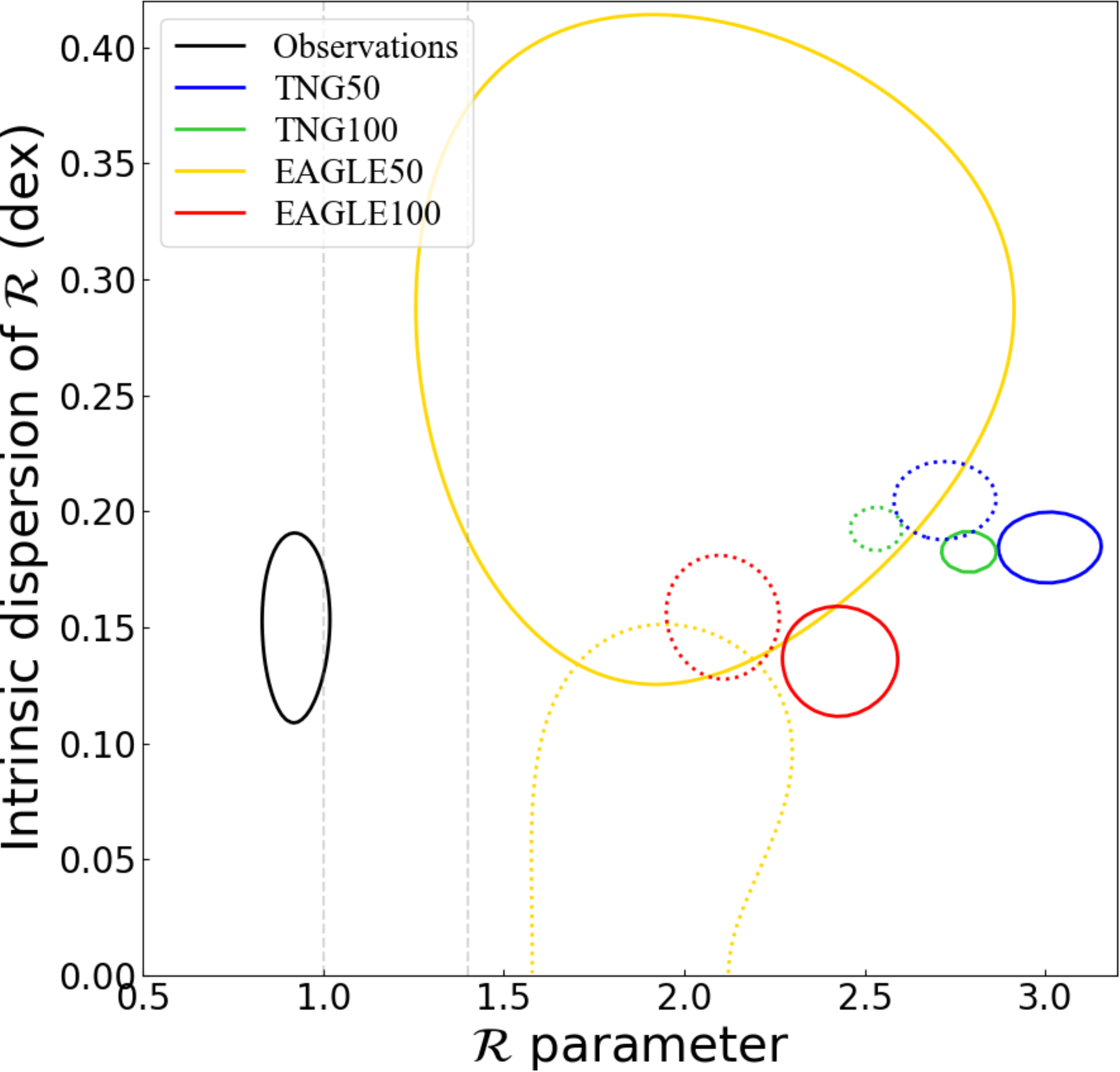}
	\caption{The posterior inference on the mean value of $\mathcal{R}$ and its intrinsic dispersion, both calculated in log-space (i.e., the $x$-axis shows $10^{\overline{\mathcal{\widetilde{R}}}}$, see the text). The vertical dashed grey lines demarcate the fast bar regime. The solid black (green, blue, red, yellow) contour corresponds to $1\sigma$ outliers from the observed (TNG100, TNG50, EAGLE100, EAGLE50) posterior. The corresponding dotted curves are based on $R_{\text{CR}}$ measurements that use $v_{\phi}^{\text{rms}}$ for the rotation curve (see Section \ref{newCR} for more details). Note that with different methods for measuring the corotation radius, different galaxies in each simulation pass the quality cuts.}
	\label{fig6}
\end{figure}

Fig. \ref{fig6} shows our posteriors on $\left(\overline{\mathcal{\widetilde{R}}}, \sigma_{\mathcal{\widetilde{R}}} \right)$ based on a high-resolution grid in both parameters, with the resulting array then normalized to a sum of 1. There is a very significant mismatch between the TNG50 and observational posteriors, mainly because observations prefer $\mathcal{R} \approx 1$ while TNG50 galaxies prefer $\mathcal{R} \approx 3$ with slightly more scatter $-$ this is also evident from Tables \ref{Obs_summary_statistics} and \ref{TNG_summary_statistics}.

To quantify the probability that e.g. TNG50 galaxy bars are compatible with observations, we find the tension between each $\left(\overline{\mathcal{\widetilde{R}}}, \sigma_{\mathcal{\widetilde{R}}} \right)$ pair and the simulated distribution. The tension is calculated as the weighted average over all these determinations, with each $\left(\overline{\mathcal{\widetilde{R}}}, \sigma_{\mathcal{\widetilde{R}}} \right)$ weighted by the observational likelihood that it is correct. We then convert the so-obtained probability into an equivalent number of standard deviations for a Gaussian random variable. In this way, we find that the TNG50 results are incompatible with observations at $12.62\, \sigma$ confidence, demonstrating the seriousness of the problem given also the excellent resolution of this simulation (Section \ref{TNG}). The tension with the here-considered $\Lambda$CDM simulations is summarized in Table \ref{Tension_summary}.

\citet{Kim_2021} have recently noted that bar lengths normalized by the disc scale length do not show any strong cosmic evolution over the period $0.2 < z \leq 0.835$, and that this result is consistent with up-to-date simulations in the $\Lambda$CDM context. This agreement is independent of our result that the simulated bar lengths at low redshift are too small for the corotation radius given by the pattern speed. Indeed, if dynamical friction with the CDM halo slows the bar down while normalized bar lengths do not evolve much, one ends up with slow bars in terms of the ratio between corotation radius and bar length. This is indeed the case in up-to-date simulations as shown here, but disagrees with observations.

\begin{table}
	\centering
	\begin{tabular}{lcc}
		\hline
        & \multicolumn{2}{c}{Rotation curve method} \\
		Simulation & $v_c$ & $v_{\phi}^{\text{rms}}$ \\ \hline 
		TNG100 & $13.56\sigma$ & $12.06\sigma$ \\
		TNG50 & $12.62\sigma$ & $11.11\sigma$ \\
		EAGLE100 & $9.69\sigma$ & $8.14\sigma$ \\
		EAGLE50 & $2.94\sigma$ & $2.68\sigma$ \\
		\hline
	\end{tabular}
	\caption{The level of tension between the observed $\mathcal{R}$ parameter distribution and that in different $\Lambda$CDM cosmological simulations, shown for two methods of calculating the rotation curve (Section \ref{CR}). The low tension in EAGLE50 is due to a very small sample size of 5 (Table \ref{Sample_sizes}), but its $\mathcal{R}$ parameter distribution is consistent with the other simulations (Fig. \ref{fig6}).}
	\label{Tension_summary}
\end{table}

\subsection{Corotation radius measured by $v_{\phi}^{\text{rms}}$}
\label{newCR}

As mentioned in Section \ref{CR}, another choice for the rotation curve is $v_{\phi}^{\text{rms}}$. Though unsuitable for a comparison with observations, it is useful in the sense that it underestimates the rotation curve, thereby reducing the bar speed tension. If the tension remains strong even when using $v_{\phi}^{\text{rms}}$, this would guarantee the existence of the discrepancy. It turns out that this is indeed the case $-$ even with this determination of the rotation curve, the discrepancy remains dramatic. More specifically, the average value of $\mathcal{R}$ is now $\overline{\mathcal{R}}=2.52$ (TNG100), $2.74$ (TNG50), $2.08$ (EAGLE100), and $1.98$ (EAGLE50). The $1\sigma$ contours are shown by dotted curves in Fig. \ref{fig6}. As expected, a tangible shift to the left is clearly seen compared to the solid contours of the same colour, which show results using our nominal rotation curve determinations. Consequently, the level of tension in TNG100, TNG50, EAGLE100, and EAGLE50 reduces to $12.06\sigma$, $11.11\sigma$, $8.14\sigma$, and $2.68\sigma$, respectively (Table \ref{Tension_summary}). It is therefore clear that the tension cannot be alleviated by any plausible choice of method to find the rotation curve. We stress that in all cases, the tension is even larger in our nominal analysis, where we adopt $v_c$ for the rotation curve due to the higher accuracy (Fig. \ref{RC}).

\section{Discussion}
\label{Discussion}

In all four considered $\Lambda$CDM cosmological simulations (EAGLE50, EAGLE100, TNG50, and TNG100), most galaxy bars are slow in the sense of being much shorter than their corotation radius ($\mathcal{R} \gg 1.4$). However, the majority of observed galaxies have fast bars \citep{Cuomo2020}. The average observed value of $\mathcal{R}$ is around 1, whereas in cosmological simulations it exceeds 1.9. It is striking that in the most recent and highest resolution simulation (TNG50), the average value of $\mathcal{R}$ has the largest value of $\approx 3.0$ among the here-considered simulations.\footnote{The overall tension is highest in TNG100 because it has a larger sample size than TNG50.}

To quantify the tension between standard cosmological simulations and galactic bar pattern speed observations, we assumed that galaxies obey a log-normal distribution of $\mathcal{R}$ with an unknown mean and intrinsic dispersion. The observationally inferred values of the two population parameters differ from that in TNG50 at $12.62\sigma$ significance. For the other simulations, the level of tension is $13.56\sigma$ (TNG100), $2.94\sigma$ (EAGLE50), and $9.69\sigma$ (EAGLE100). The low tension with EAGLE50 is due to the fact that it only has 6 galaxies that satisfy all our criteria to have a reliable pattern speed, greatly increasing the uncertainties.\footnote{Only 5 of the 6 EAGLE50 galaxies have a sufficiently precise $\mathcal{R}$ to be usable in our statistical analysis.} Our reported $9.69\sigma$ tension for EAGLE100 is similar to the $7.96\sigma$ tension estimated previously by \citet{Roshan_2021}, which was based on 48 $\mathcal{R}$ parameter values obtained from \citet{Algorry_2017}. Since the statistical analysis applied in \citet{Roshan_2021} is identical to that applied here, this approximate agreement in the level of tension suggests that our $\mathcal{R}$ parameter determinations are broadly in line with those of other workers.

Using a lower estimate for the rotation curve amplitude does not change the story significantly. In this case, we showed in Section \ref{newCR} that the discrepancy is reduced to $11.11\sigma$, $12.06\sigma$, $2.68\sigma$, and $8.14\sigma$ for the TNG50, TNG100, EAGLE50, and EAGLE100 run, respectively. Another issue is that a wrong estimate of the bar length may have corrupted the estimate of $\mathcal{R}$, as pointed out for a small subsample of real galaxies with apparently ultrafast bars \citep{Cuomo2021}. Since we adopted here the same methods to recover the bar length in simulated galaxies, an erroneous measurement of some bar lengths, if present, should affect both real and simulated galaxies in the same way.

\begin{figure}
	\centering
	\includegraphics[width=0.45\textwidth]{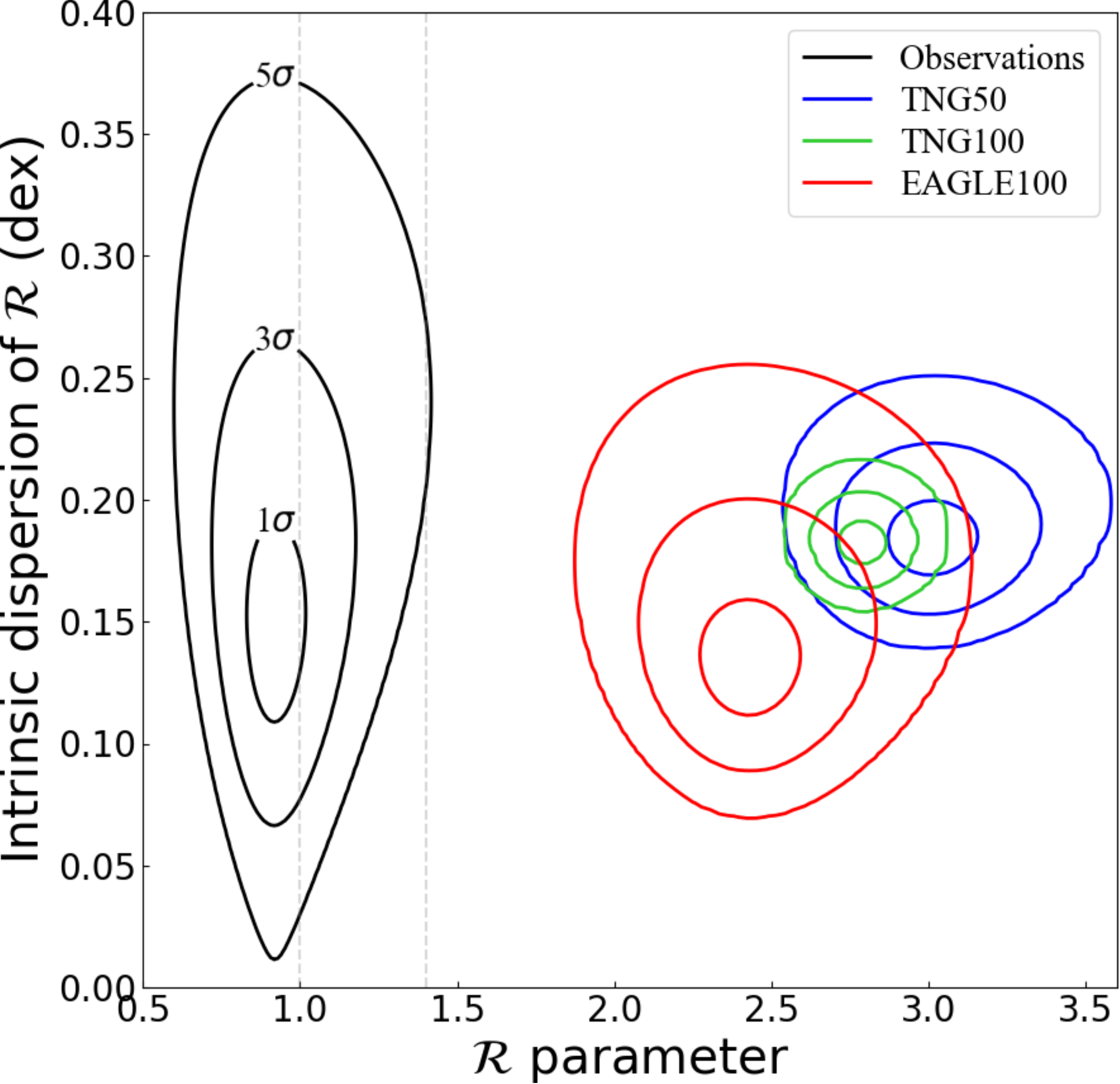}
	\caption{The posterior inference on the population logarithmic mean $\mathcal{R}$ and its intrinsic dispersion. The vertical dashed grey lines demarcate the fast bar regime. The black (green, blue, red) contours correspond to $1\sigma$, $3\sigma$, and $5\sigma$ outliers from the observed (TNG100, TNG50, EAGLE100) posterior. There are only 6 galaxies in the EAGLE50 run with known $\mathcal{R}$, so the relevant contours are too wide to be captured in this figure. Nonetheless, its $1\sigma$ contour is shown in Fig. \ref{fig6}. Notice that all four simulations considered here give compatible results with each other, but the overlap region is not consistent with observations.}
	\label{fig7}
\end{figure}

The different levels of tension with observations in different simulations can largely be attributed to sample size variations, since the statistical properties of galaxy bars are consistent between the different simulations we consider (Fig. \ref{fig7}). Thus, the fact that observed bars are typically fast remains challenging for the latest \emph{large-box} $\Lambda$CDM cosmological simulations. It is also a challenge for the zoom-in simulations of \citet{Zana_2018, Zana_2019}.

We however note that a recent study explored the 16 most strongly barred galaxies among the small sample of 30 Auriga zoom-in galaxy simulations \citep{Grand_2017}, and found that these simulated bars do stay fast \citep{Fragkoudi_2021}. However, their sample is small by construction, which is why large-box simulations are generally preferred to draw conclusions regarding overall galaxy statistics in $\Lambda$CDM. Having some fast bars is to be expected (Fig. \ref{fig5}), and a small sample is more likely to be biased and to evade other observational constraints on the overall galaxy statistics. Understanding the reasons for these different results is nevertheless of course highly interesting: \citet{Fragkoudi_2021} state that resolution is likely not the main culprit, which we agree with as TNG50 reaches almost the same resolution but typically produces slow bars. \citet{Fragkoudi_2021} instead attribute their success to their simulated disc galaxies being much less dark matter dominated than in large-box simulations, and than expected from abundance matching. It is unclear whether this would give consistent results on the overall galaxy statistics like the luminosity function in a large-box simulation. Furthermore, their findings do not extend to lower mass galaxies compared to the lowest masses in \citet{Cuomo2021}: invoking similarly low dark matter fractions in low mass galaxies would lead to conflict with observations \citep[as discussed in][]{Roshan_2021}. Finally, we note that while the Auriga simulations manage to mostly avoid the formation of classical bulges (which probably are part of the problem for both the fraction and pattern speeds of simulated bars in large-box simulations), this success is at the expense of rather unrealistically massive stellar haloes \citep[see e.g.][]{Peebles_2020}.

The slow bars predicted by $\Lambda$CDM are a consequence of dynamical friction from the surrounding dark halo particles, as shown explicitly by e.g. \citet{Roshan_2021}, where fast bars were produced by using a rigid halo that would boost the rotation curve in a similar way to a live halo, but would not cause dynamical friction on the disc. Therefore, our results could imply that dark matter should possess specific features that suppress the dynamical friction on galaxy scales, by having a behaviour very different from a sea of gravitating particles. This could be achieved by ultralight bosons \citep{Hui_2016}, but it has recently been shown that these particles are inconsistent with Lyman-$\alpha$ forest observations at 99.7\% confidence \citep{Rogers_2021}. Superfluid dark matter would also strongly suppress dynamical friction on galactic bars \citep{Berezhiani_2019}, but it was pointed out in section 5.6 of \citet{Roshan_2021} that this could be incompatible with the observed Local Group satellite planes \citep{Pawlowski_2018}. This is because some of their member satellites would be outside the superfluid halo of the parent galaxy, causing the self-gravity of the satellite's baryons to receive no further superfluid enhancement. Another problem is that stars on circular orbits would often be moving faster than the local sound speed of the superfluid, leading to the emission of Cherenkov radiation from the star that would lead to orbital decay in a small fraction of a Hubble time \citep{Mistele_2021}.

As our final remark in this discussion, let us mention that the significant failure of $\Lambda$CDM cosmological simulations with regards to bar pattern speeds is in line with several previously documented significant failures of $\Lambda$CDM in other respects \citep[e.g.][]{Kroupa_2012, Kroupa_2015, Pawlowski_2018}, including also on cosmological scales \citep[e.g.][]{Haslbauer_2020, Asencio_2021, Valentino_2021}. Taken in combination, these failures cast doubt on the existence of particle dark matter haloes around galaxies, with consequent implications for our understanding of gravity. In this regard, it is worth mentioning that in extended gravity models where CDM particles are not present around galaxies, there is obviously no dynamical friction from the halo. Therefore, such models predict fast galactic bars in a natural way \citep[e.g.,][]{Roshan_2021}. These theories are also expected to yield a rather different evolution for the bar strength, which may better explain the observed properties of M33 \citep{Sellwood_2019, Banik_2020_M33}. However, there is no sufficiently high resolution cosmological simulation in the context of extended gravity theories \citep[though lower resolution cosmological simulations exist, see e.g.][]{Katz_2013}. Consequently, it is not yet possible to fairly compare them with the standard cosmological model concerning issues like the fraction and pattern speeds of bars in disc galaxies.

\section{Conclusions}
\label{Conclusions}

In this paper, we investigated the bar pattern speeds of disc galaxies in the standard cosmological simulations known as EAGLE and IllustrisTNG. To mimic the procedure used by observers on real galaxies, we used only the final simulation snapshot and analysed it with the \citetalias{Tremaine1984} method. We showed that these simulations significantly overpredict the $\mathcal{R}$ parameter, which is the ratio of corotation radius to bar semi-major axis (Equation \ref{R_definition}). The tension is highly significant (Table \ref{Tension_summary}), and reaches $13.56\sigma$ in TNG100. In the more recent and higher resolution TNG50, the tension is $12.62\sigma$, which we suggest is largely due to a smaller sample size. Therefore, the latest cosmological $\Lambda$CDM simulations predict bars which are much slower than observed bars.

In addition to this significant mismatch in pattern speeds, $\Lambda$CDM simulations also yield a bar fraction rather different to what is inferred observationally (Fig. \ref{fig1}). This result seems robust because the bar fraction is quite similar between TNG50 and TNG100 in different stellar mass bins, though EAGLE100 has a somewhat lower bar fraction. 

In conclusion, it is clear that the latest $\Lambda$CDM simulations do not yield realistic properties for galaxy bars.

\section*{Data availability}

The simulation data used in this paper are publicly available on the EAGLE and IllustrisTNG websites (see Section \ref{cosim} for the relevant references and websites). The pattern speed observations used in this paper are available in \citet{Cuomo2020}. All the data produced in this paper (e.g. pattern speeds, corotation radii, bar lengths etc.), are available upon request to MR.

\section*{Acknowledgements}

IB was supported by a `Pathways to Research' fellowship from the University of Bonn. He acknowledges support from an Alexander von Humboldt Foundation postdoctoral research fellowship (2018-2020). BF acknowledges funding from the Agence Nationale de la Recherche (ANR projects ANR-18-CE31-0006 and ANR-19-CE31-0017), and from the European Research Council (ERC) under the European Union's Horizon 2020 Framework programme (grant agreement number 834148). VC acknowledges support from the European Southern Observatory-Government of Chile Joint Committee programme ORP060/19. The main part of the calculation was conducted at the Sci-HPC centre of the Ferdowsi University of Mashhad. The IllustrisTNG simulations were undertaken with computer time awarded by the Gauss Centre for Supercomputing (GCS) under GCS Large-Scale Projects GCS-ILLU and GCS-DWAR on the GCS share of the supercomputer Hazel Hen at the High Performance Computing Center Stuttgart (HLRS), as well as on the machines of the Max Planck Computing and Data Facility (MPCDF) in Garching, Germany. MR would like to thank Habib Khosroshahi for useful comments. We would also like to thank Annalisa Pillepich and Francesca Fragkoudi for insightful comments. The authors are grateful to the referee for comments which helped to clarify this publication and make its conclusions more robust.

\bibliographystyle{mnras}
\bibliography{short,fast_bar_tension_rev2} 

\bsp
\label{lastpage}
\end{document}